\documentclass[a4paper,11pt]{article}
\pdfoutput=1
\usepackage{jcappub}
\usepackage{epsfig}
\usepackage{placeins}
\usepackage{amssymb}
\usepackage{color}
\usepackage{float}
\usepackage{hyperref}
\usepackage{yfonts}
\usepackage{amsmath}

\newcommand{\bk}{{\mathbf k}}

\newcommand{\bx}{{\mathbf x}}
\newcommand{\bell}{{\boldsymbol \ell}}

\newcommand{\bn}{{\mathbf n}}

\newcommand{\Om}{\Omega}

\newcommand{\be}{\begin{equation}}
\newcommand{\ee}{\end{equation}}

\newcommand{\bea}{\begin{eqnarray}}
\newcommand{\eea}{\end{eqnarray}}
\newcommand{\bean}{\begin{eqnarray*}}
\newcommand{\eean}{\end{eqnarray*}}
\newcommand{\dd}{\partial}

\newcommand{\HH}{{\cal H}}
\newcommand{\RR}{{\cal R}}

\newcommand{\bxx}{{\boldsymbol \Delta \bx}}
    \newcommand{\ndv}{\bn \cdot {\bf v}}

\usepackage{color}

\begin{document}

\title{Lensing smoothing of BAO wiggles}

\author[a, b]{Enea Di Dio}
\emailAdd{enea.didio@oats.inaf.it}
\affiliation[a]{INAF - Osservatorio Astronomico di Trieste, Via G. B. Tiepolo 11, I-34143 Trieste, Italy}
\affiliation[b]{INFN, Sezione di Trieste, Via Valerio 2, I-34127 Trieste, Italy}

\date{\today}

\abstract{
We study non-perturbatively the effect of the deflection angle on the BAO wiggles of the matter power spectrum in real space. We show that from redshift $z\sim2$ this introduces a dispersion of roughly $1 \ \text{Mpc}$ at BAO scale, which corresponds approximately to a $1\%$ effect. The lensing effect induced by the deflection angle, which is completely geometrical and survey independent, smears out the BAO wiggles. The effect on the power spectrum amplitude at BAO scale is about $0.1 \%$ for $z\sim2$ and $0.2 \%$ for $z\sim4$. We compare the smoothing effects induced by the lensing potential and non-linear structure formation, showing that the two effects become comparable at $z \sim 4$, while the lensing effect dominates for sources at higher redshifts. We note that this effect is not accounted through BAO reconstruction techniques.
}

\maketitle

%%%%%%%%%%%%%%%%%%%%%%%%%
\section{Introduction}
\label{sec:intro}
All cosmological observations are performed by detecting photons that have traveled along our past light cone. Therefore they do not only carry information about the density and the velocity fields of the sources, but they are also contaminated by the geometrical effects induced by the matter distribution along the line of sight. Photons traveling in a clumpy universe are affected, to first order in perturbation theory, by cosmic magnification, integrated Sachs-Wolfe (ISW) and Shapiro time-delay effects. 
For different cosmological probes some of these effects have been already detected, see e.g.~\cite{Ade:2015zua, Ade:2015dva, Scranton:2005ci}.

Recently, motivated by the ultra-large scales that will be probed by future surveys, several works~\cite{Yoo:2009,Yoo:2010,Bonvin:2011bg,Challinor:2011bk} have derived a description of galaxy clustering in a relativistic framework. It has been shown that the leading correction to the matter power spectrum, beyond redshift space distortion, arises from the cosmic magnification, see e.g.~\cite{Scranton:2005ci} for a detection and \cite{sachs,Bartelmann:1994ye,Dolag:1997yt,Sanz:1997pw} for the first theoretical predictions. Other relativistic effects have been shown to be negligible with a single tracer analysis~\cite{Yoo:2012se, Yoo:2013zga, Alonso:2015uua}, while multi-tracer techniques~\cite{Seljak:2009} may lead to a detection with future surveys~\cite{Yoo:2012se, Bonvin:2013ogt,Alonso:2015sfa, Fonseca:2015laa, Irsic:2015nla, Gaztanaga:2015jrs,Bonvin:2015kuc}. In galaxy surveys, cosmic magnification affects the number density by changing the solid angle $d\Omega$. At the same time, since galaxy surveys are limited in flux magnitude, some faint galaxies might be enough magnified to be detected and vice-versa. The latter effect is called magnification bias and its amplitude is determined by the slope of the luminosity function. Therefore its amplitude and relevance are survey dependent. This introduces some additional systematics which have to be carefully treated. Several works have already shown its importance for (future) galaxy surveys, see e.g.~\cite{DiDio:2013sea, Montanari:2015rga, Alonso:2015sfa, Cardona:2016qxn,DiDio:2016ykq}.

It is well known that the Cosmic Microwave Background (CMB) is not affected by cosmic magnification. Indeed the change in the solid angle $d\Omega$ is exactly compensated by the change in the observed flux, due to surface brightness conservation. 
The lensing of CMB arises from the deflection angle effect, see the Ref.~\cite{Lewis:2006fu} for a detailed review about CMB lensing. This is a second order effect, but it is amplified being coherently summed along the line of sight, introducing a deflection of few arc minutes. Relativistic number counts beyond linear order have been recently computed~\cite{Yoo:2014sfa,Bertacca:2014dra,DiDio:2014lka,Nielsen:2016ldx} and they include lensing terms due to the deflection angle. These effects have been computed for the tree level bispectrum~\cite{DiDio:2015bua}. As shown in~\cite{DiDio:2015bua}, this contribution is not affected by the slope of the luminosity function (i.e.~magnification bias). At the two-point statistics, the deflection angle effect appears only at higher orders in perturbation theory.
Therefore, even if a galaxy catalog is characterized by a luminosity function that compensates the cosmic magnification contribution through the magnification bias, the deflection angle still induces a non-vanishing lensing effect on the observable.
 In this work we compute, non-perturbatively in the deflection angle, the amplitude of this effect on the Baryonic Acoustic Oscillation (BAO) wiggles of the matter power spectrum in real space. A similar calculation has been already performed for the 2-point correlation function~\cite{Dodelson:2008qc}, while in the CMB framework, an analog calculation has been first done in~\cite{Seljak:1995ve} for the angular power spectrum. Starting from the results of~\cite{Dodelson:2008qc}, we compute the lensed matter power spectrum. We show that the deflection angle smears the BAO wiggles and its effect grows with redshift, becoming larger than the smoothing effect induced by non-linear structure formation for sources at high redshift. We remark that our derivation applies also for 21cm power spectrum, see e.g.~\cite{Mandel:2005xh}. 

The paper is organized as follows. In Sec.~\ref{sec:deflaction_angle} we define and compute the linear deflection angle, while in Sec.~\ref{sec:PK} we derive the lensed matter power spectrum in real space and we show the smoothing effect on the BAO wiggles. In Sec.~\ref{sec:comparison} we compare the smoothing effect induced by the lensing potential and the non-linear dynamics of gravity. In Sec.~\ref{sec:conclusions} we draw our conclusions. 
In Appendix~\ref{app:LSS_NG} we compute the corrections induced by the non-Gaussianity of the gravitational potential at late time.
In Appendix~\ref{app:a} we present an analog derivation in terms of the redshift dependent angular power spectra, which are well-adapted to include relativistic corrections~\cite{Challinor:2011bk,Bonvin:2011bg}, and we show the amplitude of the correlation between the galaxy number counts and the lensing potential, induced at the linear level by cosmic magnification and magnification bias.

All the numerical results shown in the paper are derived with the following cosmological parameters: $h=0.704$, $\Omega_\text{cdm} =  0.226$, $\Omega_\text{b} = 0.045 $, and vanishing curvature. The primordial curvature power spectrum is characterized by $\sigma_8 = 0.81$, the pivot scale $k_\text{pivot} = 0.05 \  \text{Mpc}^{-1}$, the spectral index $n_s = 0.967$ and no running.
The matter and the lensing potential power spectra are computed with {\sc class}\footnote{\url{http://class-code.net}}~\cite{Lesgourgues:2011re,Blas:2011rf}.

%%%%%%%%%%%%%%%%%%%%%%%%%%%%%%%%%%%%%
\section{Deflection angle}
\label{sec:deflaction_angle}
Photons emitted from far away sources travel along null geodesics before being detected by an observer. Hence, if the universe is not exactly homogenous, the position $\bx$ of a galaxy has been deflected by a small amount $\delta \bx$. Since we assume the background to be statistically homogenous, it is clear that the effect due to the deflection angle appears only to second order in perturbation theory. Nevertheless, being coherently summed along the line of sight its amplitude can become comparable to some other first order effects. Because of the deflection angle, the observed density contrast $\tilde \delta$ at the apparent position $\bx$ coincides with local density perturbation $\delta$ at the physical position $\bx + \delta \bx$, i.e.~$\tilde \delta( \bx )= \delta ( \bx + \delta \bx )$. To determine $\delta \bx$ we need to compute the deviation of the photon geodesic solutions with respect to the background. We assume a spatially flat space time, and we work in Newtonian gauge $d\tilde s^2 = a^2 ds^2$ where
\be
ds^2 =  - \left( 1+ 2 \Psi \right) dt^2 + \left(1 - 2 \Phi \right) d\bx^2 
\ee
and $a(t)$ is the scale factor of the universe and $t$ the conformal time. The geometric perturbations $\Psi$ and $\Phi$ denote the Bardeen potentials. Since light-like geodesics are conformally invariant, the metrics $d\tilde s^2$ and $ds^2$ have the same light-like geodesics. 
We normalize the affine parameter $\lambda$ of the null geodesic such that at the light propagation 4-vector is defined by
\be
\left( n^\mu \right) = \left( \frac{dx^\mu}{d\lambda}\right) = \left( 1+ \delta n^0, \bn + \delta \bn \right) \, ,
\ee
with $| \bn | =1$ and $- \bn$ is the angle under which we observe the source on the sky. The perturbation $\delta n^0$ and $\delta \bn$ are determined by the linear geodesic equations
\bea \label{geodesic_start}
\frac{d \delta n^0}{d\lambda} &=& \dot \Phi - \dot \Psi + 2 \partial_r \Psi \, ,
\\
\frac{d \delta n^r}{d\lambda} &=& \partial_r \Phi - \partial_r \Psi - 2 \dot \Phi \, , \\
\frac{d \delta n^\theta}{d\lambda}+2 \frac{\delta n^\theta}{r} &=& -\frac{1}{r^2} \partial_\theta \left( \Psi + \Phi \right) \, , \\
\frac{d \delta n^\varphi}{d\lambda} + 2 \frac{\delta n^\varphi}{r}&=& -\frac{1}{r^2 \sin^2 \theta} \partial_\theta \left( \Psi + \Phi \right) \, ,
\label{geodesic_end}
\eea
where a dot denotes the partial time derivative $\partial_t$.
The geodesic equations~(\ref{geodesic_start}-\ref{geodesic_end}) are solved by
\bea
\delta r &=& \int_t^{t_o} \left( \Psi +\Phi \right) dt' = \int_0^r dr' \left( \Psi + \Phi \right)  \, ,
\\
\delta \theta &=& - \int_t^{t_o} dt' \frac{1}{r'^2}\int_{t'}^{t_o} \partial_\theta \left( \Psi +\Phi \right) dt'' 
\nonumber \\
&=& \int_t^{t_o}dt' \frac{t-t'}{\left( t_o -t\right) \left( t_o-t' \right)}  \partial_\theta \left( \Psi +\Phi \right)
= \partial_\theta \psi \, ,
\\
\delta \varphi &=&-  \int_t^{t_o} dt' \frac{1}{r'^2 \sin^2 \theta}\int_{t'}^{t_o} \partial_\varphi \left( \Psi +\Phi \right) dt'' 
\nonumber \\
&=&
\int_t^{t_o}dt' \frac{t-t'}{\left( t_o -t\right) \left( t_o-t' \right)} \frac{1}{\sin^2 \theta}  \partial_\varphi \left( \Psi +\Phi \right)
= \frac{1}{\sin^2 \theta} \partial_\varphi \psi \, ,
\eea
where we have introduced the lensing potential
\be
\psi = -  \int_0^r dr' \frac{r-r'}{r r'} \left( \Psi + \Phi \right) \, .
\ee
\\
We can define the displacements in the direction parallel, $\delta \bx_\parallel$, and perpendicular, $\delta \bx_\perp$, to the line of sight $\bn$ as follow
\bea \label{eq:deltax_parallel}
\delta \bx_\parallel &=& \delta r {\bf e}_r = \int_t^{t_o} \left( \Psi +\Phi \right) dt'  {\bf e}_r  \, ,\\
\delta \bx_\perp &=& r \delta \theta {\bf e}_\theta + r \sin \theta \delta \varphi  {\bf e}_\varphi 
= r {\boldsymbol \nabla}_\bn \psi = r^2 {\boldsymbol \nabla}_\perp \psi \, .
\label{eq:deltax_perp}
\eea
%%%%%%%%%%%%%%%%%%%%%%%%%%%%%%%%%%%%%
\section{Lensed matter power spectrum}
\label{sec:PK}
Being the lensing a pure transversal effect, in this section we compute the lensed matter power spectrum for transverse modes, 
\be \label{lensed_Pk}
\tilde P(k_\perp ) = \int e^{- i {\bf k_\perp} \cdot {\bxx_\perp}} \tilde \xi \left( \Delta x_\perp, \Delta x_\parallel \right) d^2 \Delta x_\perp d \Delta x_\parallel 
\ee
where 
\be \label{corr_function}
\tilde \xi \left( \Delta x_\perp, \Delta x_\parallel \right)= \langle \tilde  \delta \left( \bx_1  \right) \tilde \delta \left( \bx_2 \right) \rangle = \langle \delta \left( \bx_1 + \delta \bx_1 \right) \delta \left( \bx_2 + \delta \bx_2 \right) \rangle
\ee
is the lensed correlation function, $\bxx = \bx_2 - \bx_1$ and $\langle .. \rangle$ is the ensemble average. We remind that not only the pure transversal modes $\bk_\perp$ are lensed, but the effective transversal part of each Fourier mode ${\bf k}$ is lensed. Therefore the lensed power spectrum defined in Eq.~\eqref{lensed_Pk} is analogous to $\tilde P \left( k_\perp, k_\parallel \right)$ for each fixed $k_\parallel$.
As shown in Ref.~\cite{Mandel:2005xh} in context of the 21cm observable, the 3-dimensional lensed power spectrum $\tilde P(\bk)$ induces a scale-dependent anisotropic behaviour. In principle this angular dependence offers an opportunity to measure the lensing potential directly in galaxy clustering surveys. We leave this analysis to a future work, while we focus here on the smearing effect on BAO wiggles.

By computing explicitly the lensed correlation function, Eq.~(\ref{corr_function}), we find
\bea \label{corr_function2}
\tilde \xi \left( \Delta x_\perp, \Delta x_\parallel \right)&=& \frac{1}{\left( 2 \pi \right)^3} \int d^3k_1 d^3k_2  e^{i \bk_1 \cdot \bx_1 + i  \bk_2 \cdot \bx_2 } \langle e^{i \bk_1 \cdot \delta \bx_1+ i \bk_2 \cdot \delta \bx_2} \delta \left( \bk_1 \right) \delta \left( \bk_2 \right) \rangle
\nonumber \\
&=& 
\frac{1}{\left( 2 \pi \right)^3} \int d^3k_1 d^3k_2  \ e^{i \bk_1 \cdot \bx_1 + i  \bk_2 \cdot \bx_2 } \langle e^{i \bk_1 \cdot \delta \bx_1+ i \bk_2 \cdot \delta \bx_2}\rangle \langle \delta \left( \bk_1 \right) \delta \left( \bk_2 \right) \rangle
\nonumber \\
&=& \frac{1}{\left( 2 \pi \right)^3} \int d^3 k \  e^{i \bk \cdot \left( \bx_1 - \bx_2 \right)}  \langle e^{i \bk \cdot  \left( \delta \bx_1-   \delta \bx_2\right)}\rangle P \left( k \right) 
\nonumber \\
&=&
 \frac{1}{\left( 2 \pi \right)^3} \int d^3 k \  e^{- i \bk \cdot \bxx}   e^{- \langle \left[ \bk \cdot  \left( \delta \bx_1-   \delta \bx_2\right) \right]^2 \rangle/2} P \left( k \right) \, ,
\eea
where for the second equality we have assumed that the deflection angle $\delta \bx_\perp$ and the time delay $\delta \bx_\parallel$ effects are uncorrelated with the density perturbation at the source position.

\begin{figure}[htb!]
\begin{center}
\includegraphics[width=9cm]{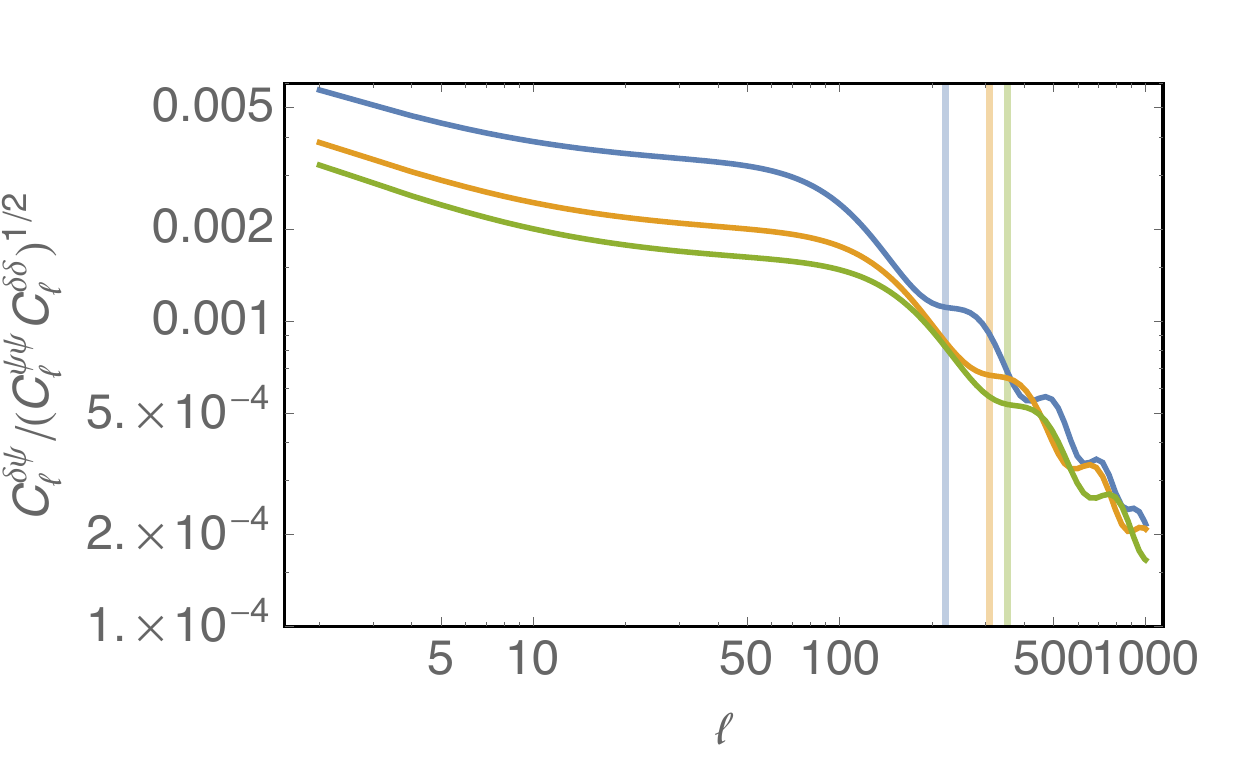}
\caption{We plot the correlation between the density fluctuation $\delta$ and the lensing potential $\psi$. Different colors refer to different redshifts: $z=2$ (blue), $z=4$ (orange), $z=6$ (green). Vertical lines show the BAO scale in $\ell$-space at the three different redshifts.}
\label{fig:delta_psi}
\end{center}
\end{figure}
As shown in Fig.~\ref{fig:delta_psi}, the density perturbation $\delta$ and the lensing potential $\psi$ are very weakly correlated. Indeed the lensing kernel peaks at half distance in comoving space between the source and the observer. Therefore for galaxies at high redshift, for which we expect the largest effect due to the lensing potential, the local density fluctuations and the metric perturbations, which source the deflection angle, are separated by hundreds or thousands of Mpc. Indeed, from Fig.~\ref{fig:delta_psi} we remark that the correlation between the lensing potential and density perturbation is smaller for higher redshifts.

Nevertheless we remark that in a full relativistic analysis, cosmic magnification and magnification bias introduce additional long range radial correlations. This correlation along the line of sight is survey dependent since its details are determined by galaxy and magnification biases. 
Even if we are interested in computing lensing correction beyond the linear order induced by cosmic magnification, and considering that we can always select a galaxy catalog such that the cosmic magnification is exactly compensated by the magnification bias, in Appendix~\ref{app:a} we show the correlation induced by cosmic magnification for different magnification bias values.
Even if the inclusion of cosmic magnification and magnification bias deserves a detailed study, we believe that this introduces a broadband effect and it does not have a relevant impact on BAO wiggles smoothing effect for current galaxy surveys. While for galaxy catalogs at higher redshift with magnification bias values which significantly deviate from\footnote{The magnification bias is defined in Eq.~\eqref{magnification_bias}. We remark that for $s=0.4$ cosmic magnification is exactly compensated by magnification bias.} $s=0.4$ the correlation between the lensing potential and the observed galaxy number counts might become relevant. 
Though, at high redshift our formalism can be applied to 21cm power spectrum, which is not affected by the long distance correlation introduced by cosmic magnification and magnification bias, see e.g.~\cite{Hall:2012wd}.

On the last line of Eq.~\eqref{corr_function2} we have also assumed that the deflection angles are described by Gaussian fluctuations. This allows us to sum up the exponential using
\be
\label{Gaussian_trick}
 \langle e^{i \bk \cdot  \left( \delta \bx_1-   \delta \bx_2\right)}\rangle  =  e^{- \langle \left[ \bk \cdot  \left( \delta \bx_1-   \delta \bx_2\right) \right]^2 \rangle/2} 
\ee
which holds for Gaussian fluctuations.
In the CMB context, the corrections, due to the non-Gaussian nature of the deflection angle, seem to be much smaller than the leading contribution coming from the non-perturbative approach we adopted here, see e.g.~\cite{Lewis:2016tuj, Marozzi:2016qxl, Marozzi:2016und}.
In Appendix~\ref{app:LSS_NG} we give an estimation of the corrections induced by the non-Gaussian gravitational potential, showing that their magnitudes are small (well below the per-mill of the power spectrum amplitude at BAO scales).
We implicitly assume also Born approximation. In CMB or shear weak lensing frameworks, it has been shown that the post-Born corrections are small, see e.g.~\cite{Krause:2009yr, Pratten:2016dsm, Marozzi:2016uob}.
 Therefore, even if we work under these assumptions, we believe that we capture the correct magnitude of the effect. A more quantitative approach would require a light-tracing analysis with N-body simulation. In particular, full relativistic N-body~\cite{Adamek:2015eda} will allow to consider both geometrical and dynamical effects together. This is well beyond the aim of our work.
We have as well assumed statistical homogeneity and isotropy and introduced the unlensed matter power spectrum $P(k)$
\be
\langle\delta \left( \bk_1\right) \delta \left( \bk_2 \right) \rangle =  \delta^{(3)}_D \left( \bk_1 + \bk_2 \right) P \left( k \right) \, .
\ee
A detailed derivation and analysis of the lensed correlation function, Eq.~(\ref{corr_function2}), has been performed in~\cite{Dodelson:2008qc}. Here we summarize the main results before computing the lensed matter power spectrum. In the Appendix~\ref{app:a} we will perform a similar derivation for the redshift dependent angular power spectra, which will not require some approximation we adopt in the rest of this section.

To compute the lensed power spectrum, we first need to compute the following expectation value
\be
\langle \left[ \bk \cdot  \left( \delta \bx_1-   \delta \bx_2\right) \right]^2 \rangle = k_i k_j \left( \langle \delta x_1^i  \delta x_1^j  \rangle + \langle \delta x_2^i  \delta x_2^j  \rangle - 2 \langle \delta x_1^i  \delta x_2^j  \rangle  \right) \, .
\ee
Following the notation of~\cite{Dodelson:2008qc}, we define the correlation matrix
\be
Z_{ij} \equiv  \frac{\langle \delta x_1^i  \delta x_1^j  \rangle + \langle \delta x_2^i  \delta x_2^j  \rangle }{2}-  \langle \delta x_1^i  \delta x_2^j  \rangle \, .
\ee
Without loss of generality we can choose the coordinates such that $\bx_1$ lies on the $z$-axis and $\bx_2$ on the $x$-$z$ plane. For sake of simplicity we denote $r=\text{min} \left( r_1, r_2 \right)$.
In this case the correlation matrix reduces to
\be
Z= \left( \begin{array} {ccc}
T_1 + T_2 + D/2    &   0 &   -V_1 \\
0       & T_1 +T_2- D/2 &   0 \\
-V_2   & 0 & S 
\end{array}
\right), 
\ee
where, under the Limber approximation~\cite{Limber:1954zz,LoVerde:2008re}, the different matrix coefficients are
\bea
&&T_1 \left( \Delta x_\parallel \right) = \Delta x_\parallel^2 \int_0^r dr' \int \frac{dk}{8 \pi} \ k^3 P_\RR \left( k \right) T^2_{\Psi + \Phi} \left( k , r' \right) \, ,
\\
&& T_2\left( \Delta x_\perp \right) = \int_0^r dr' \left( r - r' \right)^2  \int \frac{dk}{4 \pi} k^3  P_\RR \left( k \right) T^2_{\Psi + \Phi} \left( k , r' \right) \left( 1 - J_0 \left( k \Delta x_\perp \frac{r'}{r} \right) \right) \, , \\
&& D \left( \Delta x_\perp \right) = 2 \int_0^r dr' \left( r - r' \right)^2 \int \frac{dk}{4 \pi} k^3 P_\RR \left( k \right) T^2_{\Psi + \Phi} \left( k , r' \right)  J_2 \left( k \Delta x_\perp \frac{r'}{r} \right) \, , \\
&& S \left( \Delta x_\perp \right) = 2 \int_0^r dr'  \int \frac{dk}{4 \pi} k P_\RR \left( k \right) T^2_{\Psi + \Phi} \left( k , r' \right) \left( 1 - J_0 \left( k \Delta x_\perp \frac{r'}{r} \right) \right) \, ,
\\
&& \left[ V_1 +V_2 \right] \left(   \Delta x_\perp, \Delta x_\parallel \right) = 2 \Delta x_\parallel \int_0^r dr' \int \frac{dk }{4 \pi } k^2
 P_\RR \left( k \right) T^2_{\Psi + \Phi} \left( k , r' \right) J_1 \left( k \Delta x_\perp \frac{r'}{r} \right)   \, , \ \
 \eea
where $J_i$ denotes the Bessel function of order $i$.
To be consistent with the notation adopted in~\cite{Lesgourgues:2011re,Blas:2011rf,DiDio:2013bqa}, we have introduced the primordial curvature power spectrum $P_\RR \left( k \right)$ and the Bardeen transfer function $T_{\Psi + \Phi} \left( k , r \right)$ such that
 \be
 \langle \left( \Psi + \Phi \right) \left( \bk , r \right)  \left( \Psi + \Phi \right) \left( \bk' , r' \right) \rangle = \delta_D^{(3)} \left( \bk + \bk'\right) P_\RR \left( k \right) T_{\Psi + \Phi} \left( k , r \right) T_{\Psi + \Phi} \left( k' , r' \right) \, ,
 \ee
 and
 \be
 P_\RR \left( k \right) = A_s \left( \frac{k}{k_\text{pivot}} \right)^{n_s-1} \frac{2 \pi^2}{k^3} \, .
 \ee
The correlation matrix $Z$ depends on the longitudinal separation $\Delta x_\parallel$ only through the coefficients $T_1$ and $V_i$ (with $i=1,2$). Nevertheless, these terms are parametrically suppressed by a factor $\sim \left( \Delta x_\parallel /r \right)^2$ with respect to $T_2$ and $D$. This result arises from the fact the longitudinal displacement $\delta \bx_\parallel$, Eq.~\eqref{eq:deltax_parallel}, is suppressed by a spatial derivative on relevant scales with respect to the transversal displacement $\delta \bx_\perp$, Eq.~\eqref{eq:deltax_perp}. We can show that also the term $S$ is suppressed, anyway this is not relevant to compute $\tilde P \left( k_\perp \right)$ as defined in Eq.~\eqref{lensed_Pk}. Hence we will simply neglect the $\Delta x_\parallel$ dependence in the correlation matrix $Z$.

Then, by replacing Eq.~(\ref{corr_function2}) in Eq.~(\ref{lensed_Pk}), and integrating over $\Delta x_\parallel$ we obtain
\be
\tilde P\left( k_\perp \right) = \frac{1}{\left( 2 \pi\right)^2} \int d^2 \Delta x_\perp d^2 k'_\perp e^{-i \bk_\perp \cdot \bxx_\perp} e^{-i \bk'_\perp \cdot \bxx_\perp}   e^{-{k'}_\perp^\alpha {k'}_\perp^\beta Z_{\alpha \beta}} P \left( k'_\perp \right) \, 
\ee
where $\alpha$ and $\beta$ indices run over $\left\{1,2 \right\}$ and 
\be
\left( Z_{\alpha \beta}\right) \sim \left( \begin{array} {cc}
 T_2 + D/2    &   0  \\
0       & T_2 -  D/2  
\end{array}
\right).
\ee
Within this approximation we can adopt the same formalism successfully used for CMB lensing and profit from the existing Boltzmann codes {\sc class} or {\sc camb}\footnote{\url{http://camb.info}}~\cite{Lewis:1999bs}. For this purpose we define the redshift dependent angular power spectrum of the lensing potential
\be \label{lens_pot_spectrum}
C_\ell^\psi = \frac{2}{\pi} \! \int_0^{r_1} \!\!dr' \frac{r_1 - r'}{r_1 r'}\! \int_0^{r_2} \!\! dr'' \frac{r_2 - r'' }{r_2 r''}\! \int \! dk \ k^2 P_\RR \left( k \right) T_{\Psi +\Phi}\left( k , r' \right) T_{\Psi +\Phi}\left( k , r'' \right)  j_\ell \left( k r' \right) j_\ell \left( k r'' \right)   \, ,
\ee
where $j_\ell$ denotes the spherical Bessel function of order $\ell$. By applying the Limber approximation\footnote{To be consistent with the results of Ref.~\cite{Dodelson:2008qc} we do not adopt the extended Limber approximation presented in Ref.~\cite{LoVerde:2008re}. Nevertheless, as can be seen in Appendix~\ref{app:a}, the final results agree with the exact calculation in terms of angular power spectra. Here we prefer to derive the effect in terms of the commonly used 3-dimensional matter power spectrum $P\left( k \right)$.} Eq.~\eqref{lens_pot_spectrum} simplifies to
\be
C_\ell^\psi \sim \int^r_0 dr' \frac{\left( r - r' \right)^2}{r^2 r'^4} P_\RR \left( \frac{\ell}{r'} \right) T_{\Psi + \Phi}^2 \left( \frac{\ell}{r'}, r'\right) \, .
\ee
Hence, we can rewrite
\bea \label{eq:T2}
T_2 \left( \Delta x_\perp \right) &=& \frac{r^2}{2} \frac{1}{2 \pi} \int d \ell \ \ell^3 \ C_\ell^\psi  \left[ 1 - J_0 \left( \ell \frac{\Delta x_\perp}{r} \right) \right] \, ,
\\ 
\label{eq:D}
D \left( \Delta x_\perp \right) &=& r^2 \frac{1}{2 \pi} \int d \ell \ \ell^3 \ C_\ell^\psi  J_2 \left( \ell \frac{\Delta x_\perp}{r} \right) \, ,
\eea
and we obtain
\bea
k_\perp^\alpha k_\perp^\beta Z_{\alpha \beta} = k_\perp^2 \left[ T_2\left( \Delta x _\perp \right) + \cos \left( 2 \varphi \right) \frac{D \left(   \Delta x _\perp \right)}{2} \right] \, ,
\eea
where $\varphi$ denotes the polar angle of the 2-dimensional vector $\bk_\perp$.
\begin{figure}[htb!]
\begin{center}
\includegraphics[width=7cm]{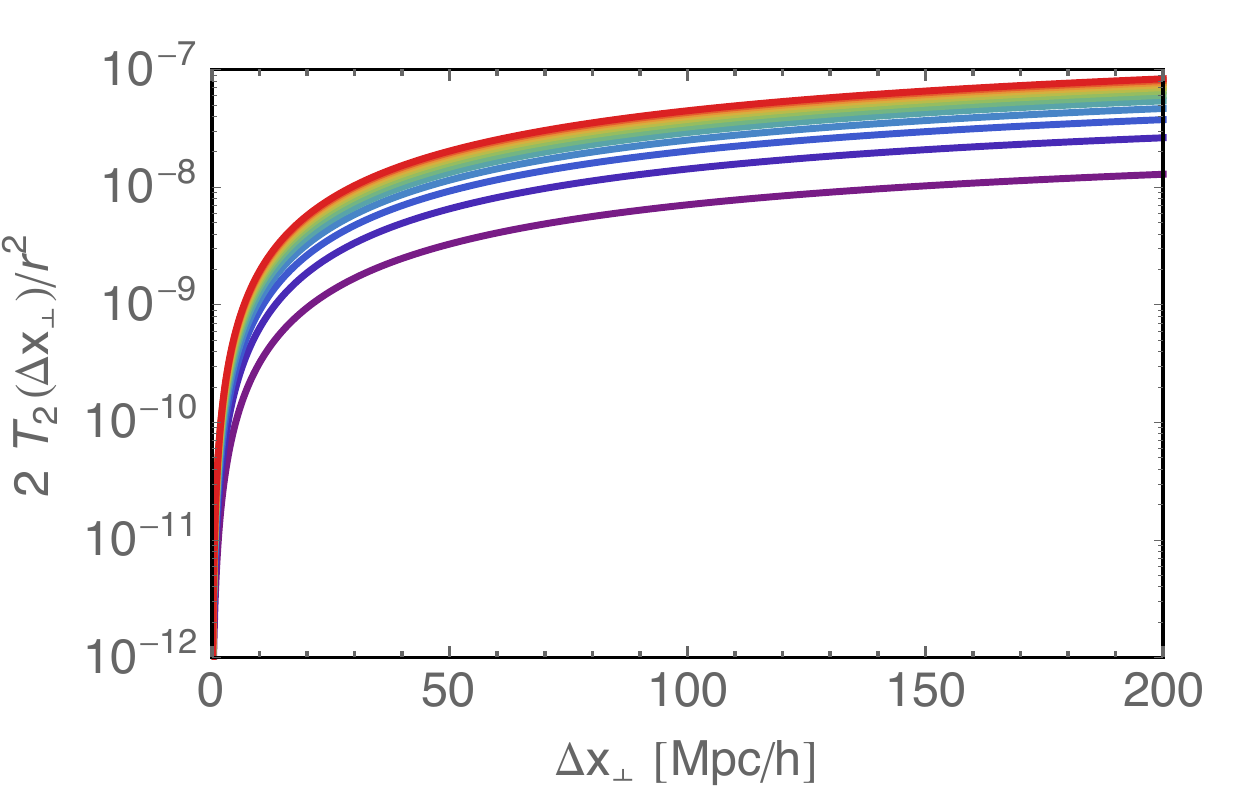}
\includegraphics[width=7cm]{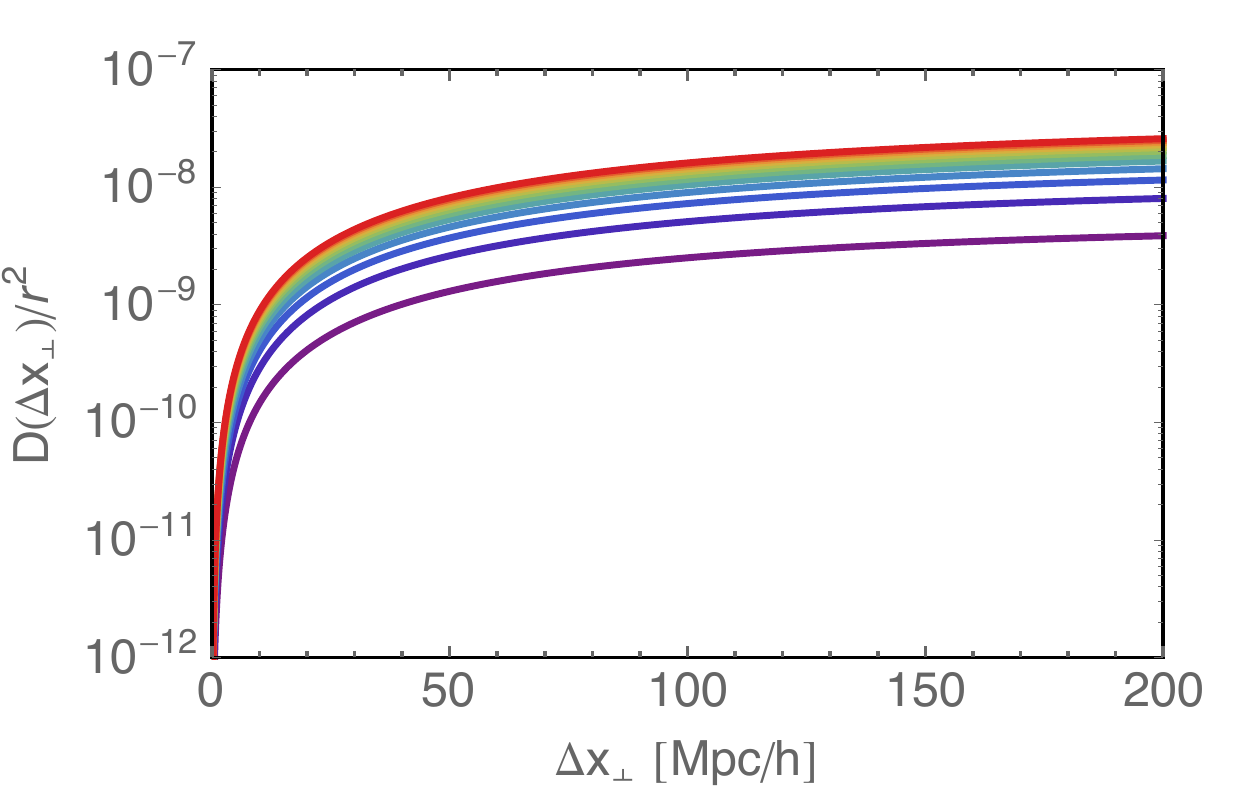}
\caption{We plot $2 T_2\left( \Delta x_\perp \right)/ r^2(z)$ (left) and $D\left( \Delta x_\perp \right)/r^2(z)$ (right) from $z=0.5$ (violet) to $z=6$ (red) with steps of $\Delta z=0.5$.}
\label{fig:T2andD}
\end{center}
\end{figure}
As we see from Fig.~\ref{fig:T2andD}, $T_2$ is always larger than $D$. So we can expand the exponential as
\bea
\exp \left(-k^\alpha k^\beta Z_{\alpha \beta} \right) &=& \exp \left( -k_\perp^2 \left[ T_2\left( \Delta x _\perp \right) + \cos \left( 2 \varphi \right) \frac{D \left(   \Delta x _\perp \right)}{2} \right]\right)
\nonumber \\
 &\sim&  e^{-k_\perp^2  T_2\left( \Delta x _\perp \right) } \left( 1 -  \cos \left( 2 \varphi \right) k^2_\perp \frac{D \left(   \Delta x _\perp \right)}{2} \right) \, .
\eea
We remind that this is an approximation on the smallness of the term $D$ with respect to $T_2$, we do not assume a small deflection angle. Finally, the lensed matter power spectrum is
\bea
\tilde P\left( k_\perp \right) &=& \frac{1}{ 2 \pi} \int d^2 \Delta x_\perp d k'_\perp \ k'_\perp  e^{-i \bk_\perp \cdot \bxx_\perp}  P \left( k'_\perp \right)e^{-{k'}_\perp^2  T_2\left( \Delta x _\perp \right) }
\nonumber \\
&& \qquad \times
 \left( 
 J_0 \left( k'_\perp \Delta x_\perp \right) 
 + {k'}_\perp^2 \frac{ D \left( \Delta x_\perp \right)}{2} J_2 \left( k'_\perp \Delta x_\perp \right)
  \right)
  \nonumber \\
  &=&
\int d \Delta x_\perp d k'_\perp \Delta x_\perp k'_\perp    P \left( k'_\perp \right)e^{-{k'}_\perp^2  T_2\left( \Delta x _\perp \right) } J_0 \left( k_\perp \Delta x_\perp \right)
\nonumber \\
&& \qquad \times
 \left( 
 J_0 \left( k'_\perp \Delta x_\perp \right) 
 + {k'}_\perp^2 \frac{ D \left( \Delta x_\perp \right)}{2} J_2 \left( k'_\perp \Delta x_\perp \right)
  \right) \, . \label{lensed_PK}
\eea
With this result we include non-perturbatively the lensing effect in the matter power spectrum. Indeed, under the assumption that the lensing potential is Gaussian, the exponential in Eq.~\eqref{lensed_PK} resums the deflection angle effects to any order in perturbation theory. Even if the physical deflection angle is not described by Gaussian statistics, due to non-linear evolution at low redshift, the amplitude of the corrections induced by the non-Gaussian gravitational potential is always well below the per-mill level of the power spectrum amplitude at BAO scales, see Appendix~\ref{app:LSS_NG}. Therefore, in particular for galaxies at high redshift, this approximation can be safely applied.

We notice that the gravitational lensing introduces a smoothing scale on the lensed matter power spectrum. It is then convenient to define the redshift dependent dispersion
\be
\sigma \left( \Delta x_\perp, z \right) \equiv \sqrt{2 \ T_2 \left( \Delta x_\perp \right)} \, .
\ee
\begin{figure}[htb!]
\begin{center}
\includegraphics[width=7cm]{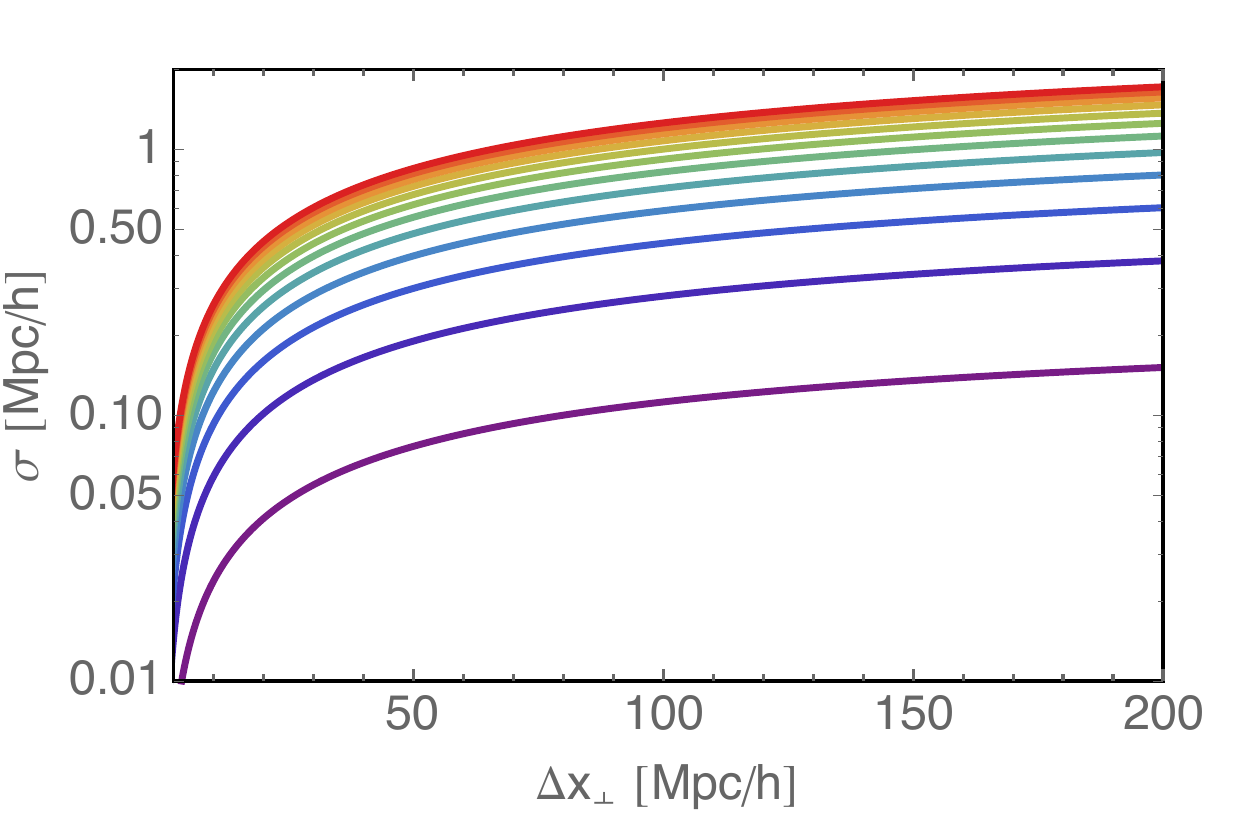}
\includegraphics[width=7cm]{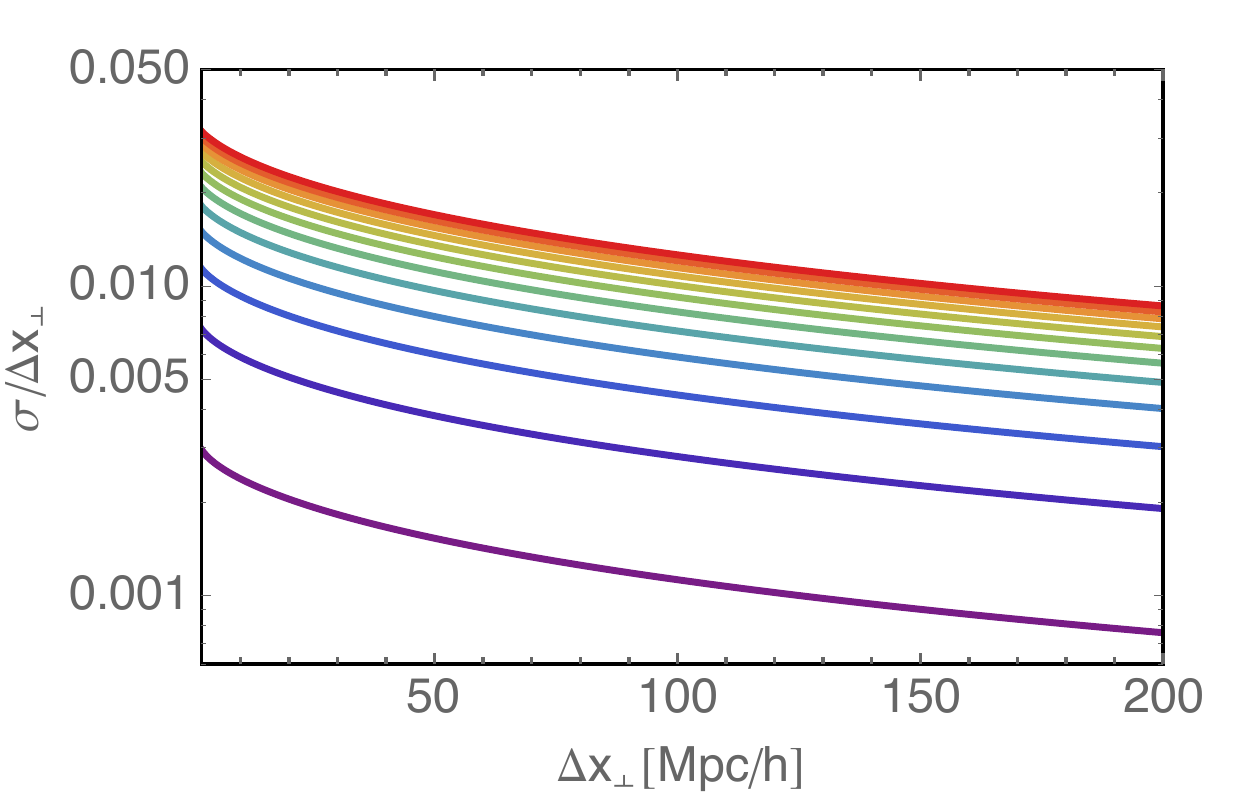}
\caption{We plot $\sigma\left( \Delta x_\perp \right)$ (left) and $\sigma \left( \Delta x_\perp \right)/ \Delta x_\perp$ (right) for $z=0.5$ (violet) to $z=6$ (red) with step-size of $\Delta z=0.5$. They show, respectively, the absolute dispersion sourced by lensing effect and the relative effect.}
\label{fig:sigma}
\end{center}
\end{figure}
As we see in Fig.~\ref{fig:sigma}, for galaxies at redshift larger than $z \sim 2$, the smearing effect produced by the lensing potential is about $1 \ \text{Mpc}$ at BAO scale. In terms of relative amplitude this translates into an effect of roughly $1\%$. This shows that, unless we measure the lensing potential independently, we have an intrinsic limit on the accuracy of measurements of transversal modes of about $1 \ \text{Mpc}$. Being the lensing effect roughly proportional to the comoving distance to the source redshift, see Eq.~\eqref{eq:T2}, the amplitude of the dispersion is not linear with the source redshift but it quickly saturates for high redshifts.

\begin{figure}[htb!]
\begin{center}
\includegraphics[width=9cm]{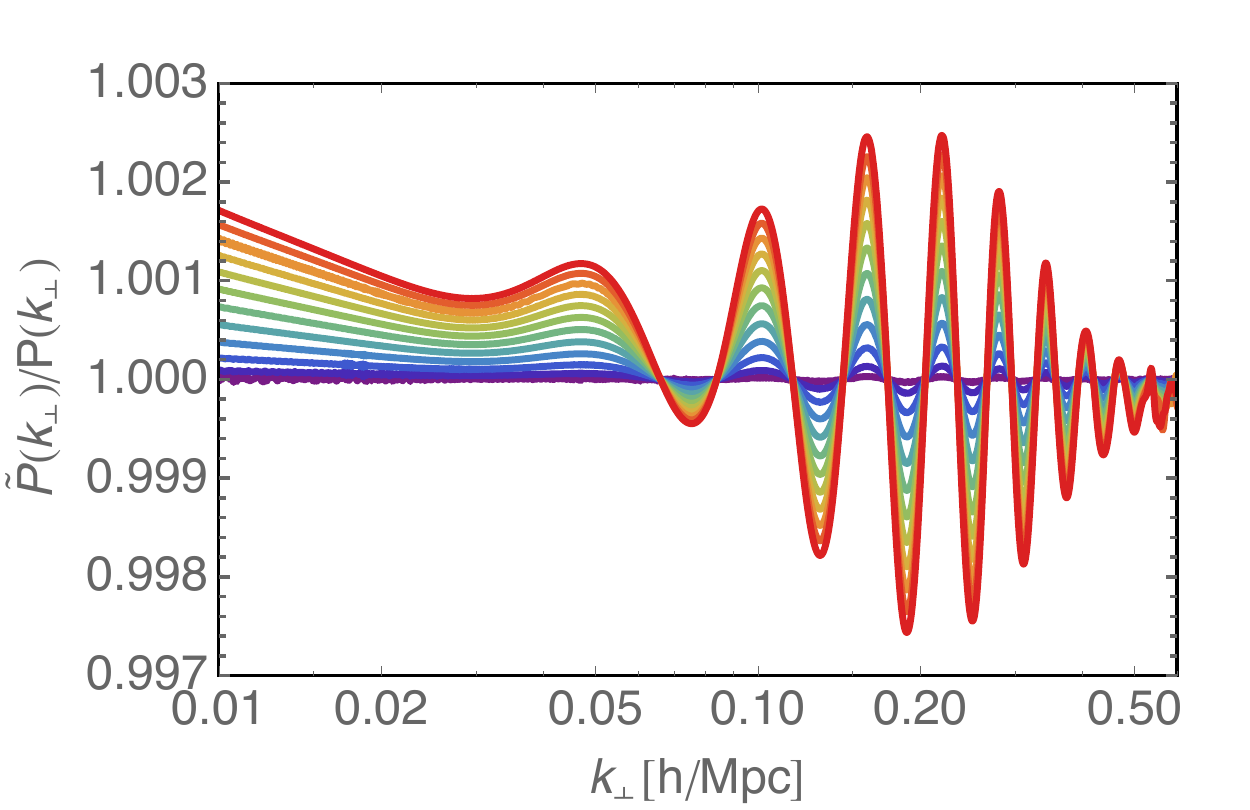}
\caption{We show the ratio between the lensed $\tilde P \left( k_\perp \right)$ and the unlensed $P \left( k_\perp \right)$ power spectra.
Different colors indicate different redshifts, from $z=0.5$ (violet) to $z=6$ (red) with step-size of $\Delta z=0.5$.
}
\label{fig:PK_ratio}
\end{center}
\end{figure}
In Fig.~\ref{fig:PK_ratio} we plot the ratio between the lensed $\tilde P( k_\perp)$ and the unlensed power spectra $P( k_\perp)$ at different redshifts. We see clearly that the effect increases with redshift reaching the amplitudes of $0.2 \% $ at $z=4$ and $0.3 \% $ at $z=6$. 

\begin{figure}[htb!]
\begin{center}
\includegraphics[width=7cm]{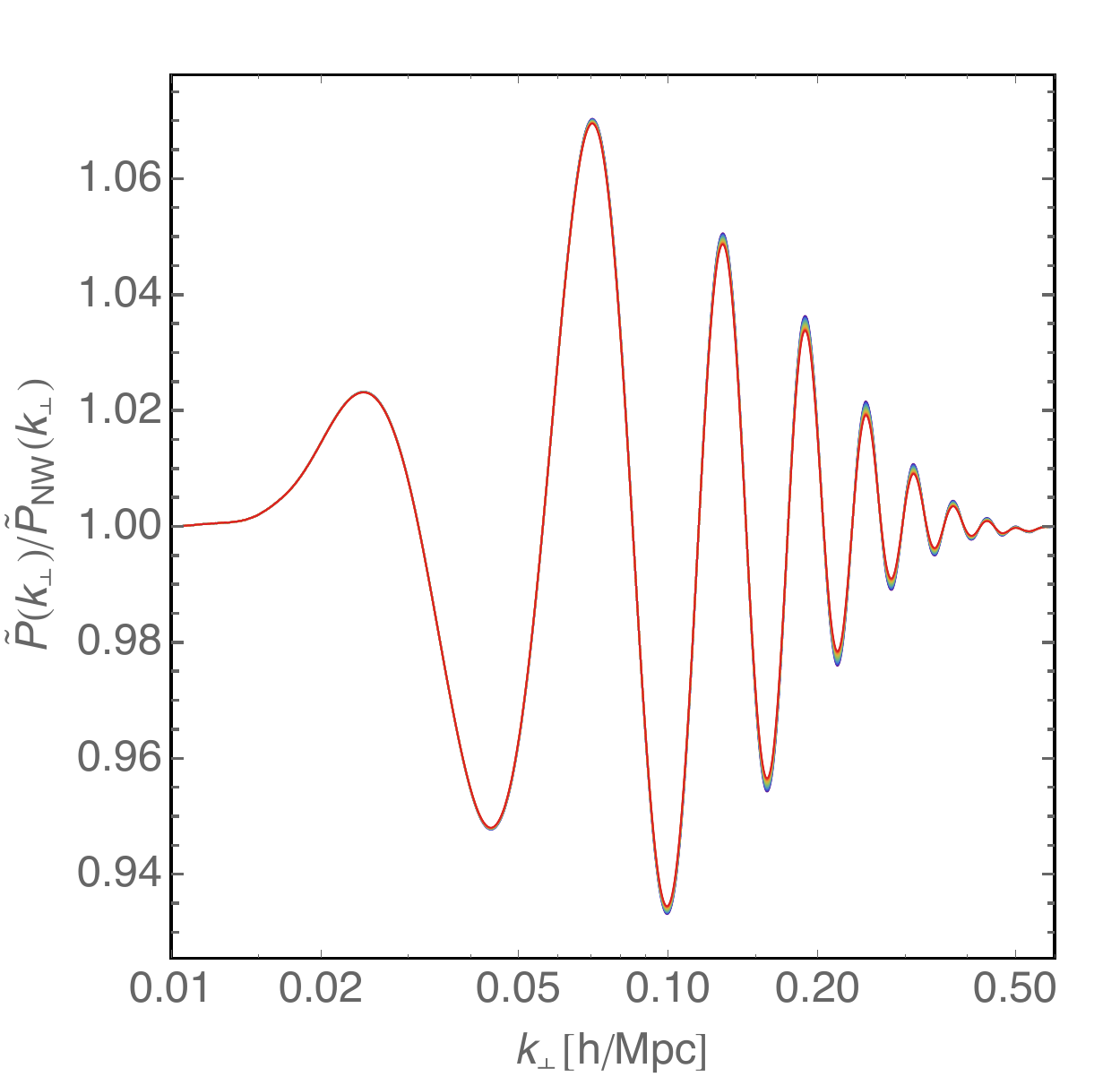}
\includegraphics[width=7cm]{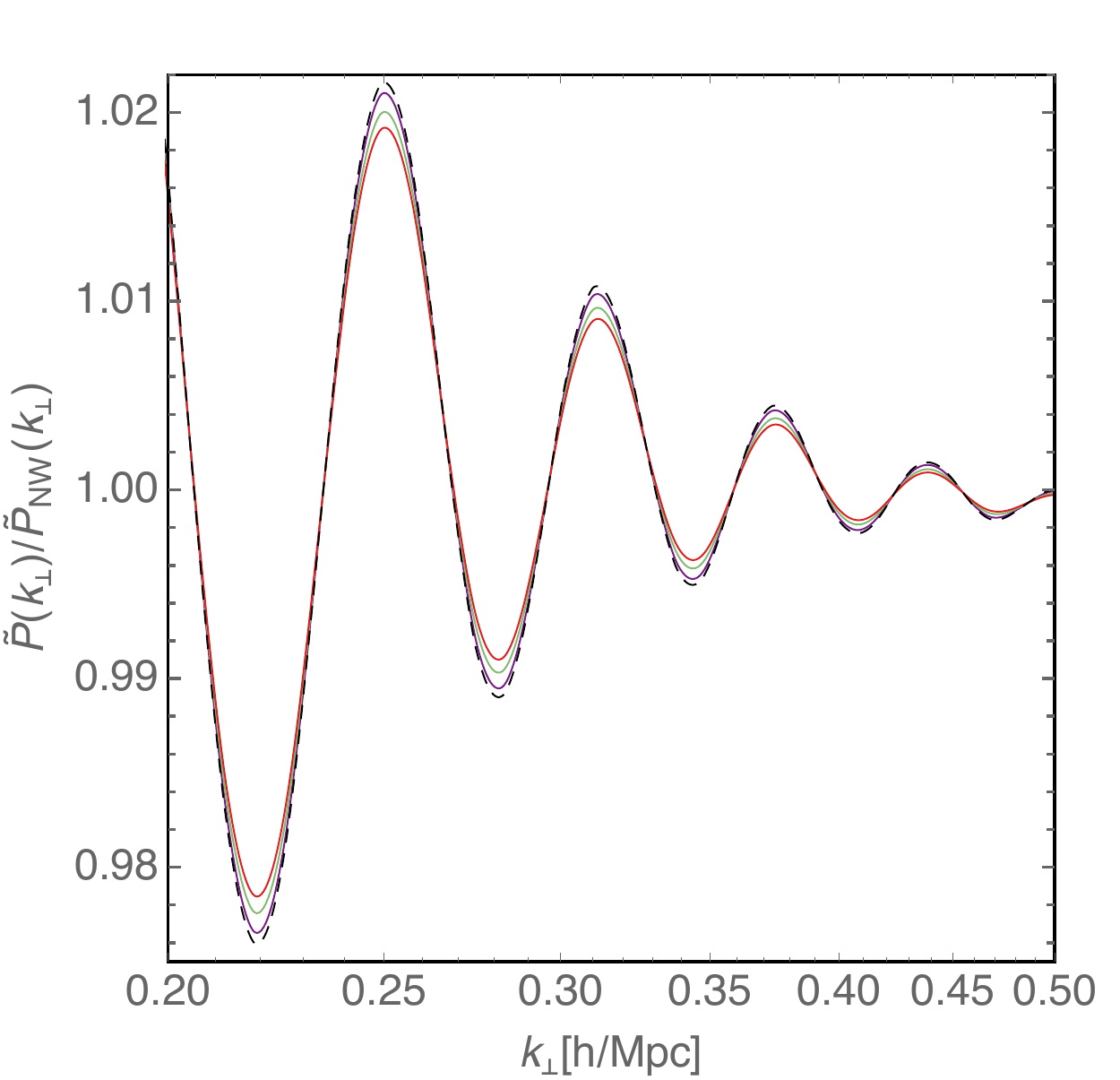}
\caption{On the left panel, we plot the ratio between the lensed power spectrum $\tilde P \left( k_\perp \right)$ and a no-wiggle lensed power spectrum $\tilde P_\text{NW} \left( k_\perp \right)$ as defined in Ref.~\cite{Eisenstein:1997ik}, and corrected to have the same amplitude at small scales as explained in details in the Appendix of Ref.~\cite{Vlah:2015zda}. Different colors indicate different redshifts, from $z=0.5$ (violet) to $z=6$ (red) with step-size of $\Delta z=0.5$. 
On the right panel we highlight the effect on the last wiggles, just showing three different redshifts: $z=2$, $z=4$ and $z=6$, together with the linear BAO wiggles (dashed black).
}
\label{fig:PK_NW}
\end{center}
\end{figure}

We notice that we have two different effects. Indeed there is a suppression induced by the exponential of the dispersion $\sigma^2$ in the lensed power spectrum, Eq.~\eqref{lensed_PK}, which manifests itself, see Fig.~\ref{fig:PK_ratio}, by the negative slope which is more relevant for sources at high redshifts. A second effect reduces the amplitude of the BAO wiggles, whose relative magnitude is determined by the amplitude of the oscillations.

In Fig.~\ref{fig:PK_NW} we highlight the effect on the BAO wiggles by dividing with an appropriate power spectrum without BAO features, see Ref.~\cite{Eisenstein:1997ik}. 
We follow the detailed procedure presented in Appendix of Ref.~\cite{Vlah:2015zda} to construct a smooth power spectrum with the same amplitude of the broadband both at small and at large scales.
Taking out the broadband suppression, the lensing effect, due to the deflection angle, smears out the BAO wiggles. The effect is larger for galaxies at high redshift. 
As wee see from the right panel of Fig.~\ref{fig:PK_NW}, the relative smearing effect is more pronounced for the wiggles at smaller scales. We remark that implicitly this smearing effect induces also a lower signal to noise ratio for the BAO wiggles measurement.

Even if the formalism adopted is based on the Gaussianity of the lensing potential, we have also checked that the difference induced by a non-linear lensing potential is very small. We have obtained the non-linear lensing potential with {\sc halofit}~\cite{Smith:2002dz,Takahashi:2012em} through {\sc class}~\cite{Lesgourgues:2011re,Blas:2011rf} code. Moreover, as in the CMB context, see e.g.~\cite{Lewis:2016tuj,Marozzi:2016qxl,Marozzi:2016und}, the correction due to the non-Gaussian nature of the deflection angle leads to a very small error, as shown in Appendix~\ref{app:LSS_NG}.

%%%%%%%%%%%%%%%%%%
\section{Comparison with non-linear smoothing}
\label{sec:comparison}
It is well known that non-linear dynamics of galaxy clustering smear out the BAO wiggles in the power spectrum. This is mainly due to the motion of galaxies which reduces the amplitude of the BAO peak in the correlation function compared to the linear evolution. In the last decade different theoretical approaches~\cite{Crocce:2005xz,Crocce:2005xy,McDonald:2006hf,Crocce:2007dt,Matsubara:2007wj,Matarrese:2007wc,Baumann:2010tm,Blas:2016sfa} have been proposed to describe the non-linear effects on BAO scales by partially resumming the non-linear contributions. Contrarily to the smoothing effect due to lensing potential, non-linearities are much more important at low redshift. Hence the two effects are relevant in different regimes. In this section we want to study in details the comparison between these two effects which lead to a smoothing of the BAO wiggles.

As shown in~\cite{Vlah:2015zda}, Zeldovich (1LPT) approximation can reproduce with good accuracy the smoothing effects on BAO wiggles compared to N-Body simulation. Even if different higher order perturbation theories agree with simulations with a better accuracy, for the purpose of this comparison we consider Zeldovich approximation only.
We follow the notation presented in~\cite{Vlah:2014nta} and we compute the matter power spectrum
\be \label{zel}
P_\text{zel}\left( k \right) = 2 \pi \int dq q^2 \int_{-1}^1 d\mu \ e^{i k q \mu} \left( e^{-\frac{1}{2} k^2 \left( X + \mu^2 Y \right)}
- e^{-\frac{1}{2} k^2 \sigma^2}  
\right) \, ,
\ee
where 
\bea
X\left( q \right) &=& \int \frac{dk}{\pi^2} P \left( k \right) \left[ \frac{1}{3} - \frac{j_1 \left( k q \right)}{k q} \right] \, , \\
Y \left( q \right) &=& \int \frac{dk}{\pi^2} P \left( k \right) j_2 \left( k q \right)\, ,
\eea
where the last term of Eq.~\eqref{zel} has been added to cure the oscillation in the integration. This term can be obtained from $X\left(q \rightarrow \infty \right)$. We apply Eq.~\eqref{zel} to compute both the matter power spectrum $P\left( k \right)$ and the no-wiggle power spectrum $P_\text{NW}\left( k \right)$.
The integral over $\mu$ is strongly oscillating, we therefore prefer to use the following identity
\be
\int d \mu e^{i A \mu} e^{B \mu^2} = 2 e^B \sum_{n =0} \left( - \frac{2 B}{A} \right)^n j_n \left( A \right) \, .
\ee
Hence, the two-dimensional integral of Eq.~\eqref{zel} simplifies to
\be
P_\text{zel} = 4 \pi \int dq q^2 \left( e^{-\frac{1}{2} k^2 \left( X + Y \right)} \sum_{n=0} \left( \frac{ k Y}{q} \right)^n j_n \left( k q \right)  - e^{- \frac{1}{2} k^2 \sigma^2} j_0 \left( k q\right) \right) \, .
\ee
The sum over $n$ converges quickly. Here we are very conservative and we consider $n=25$, as indicated by Ref.~\cite{Vlah:2014nta}.

\begin{figure}[htb!]
\begin{center}
\includegraphics[width=7cm]{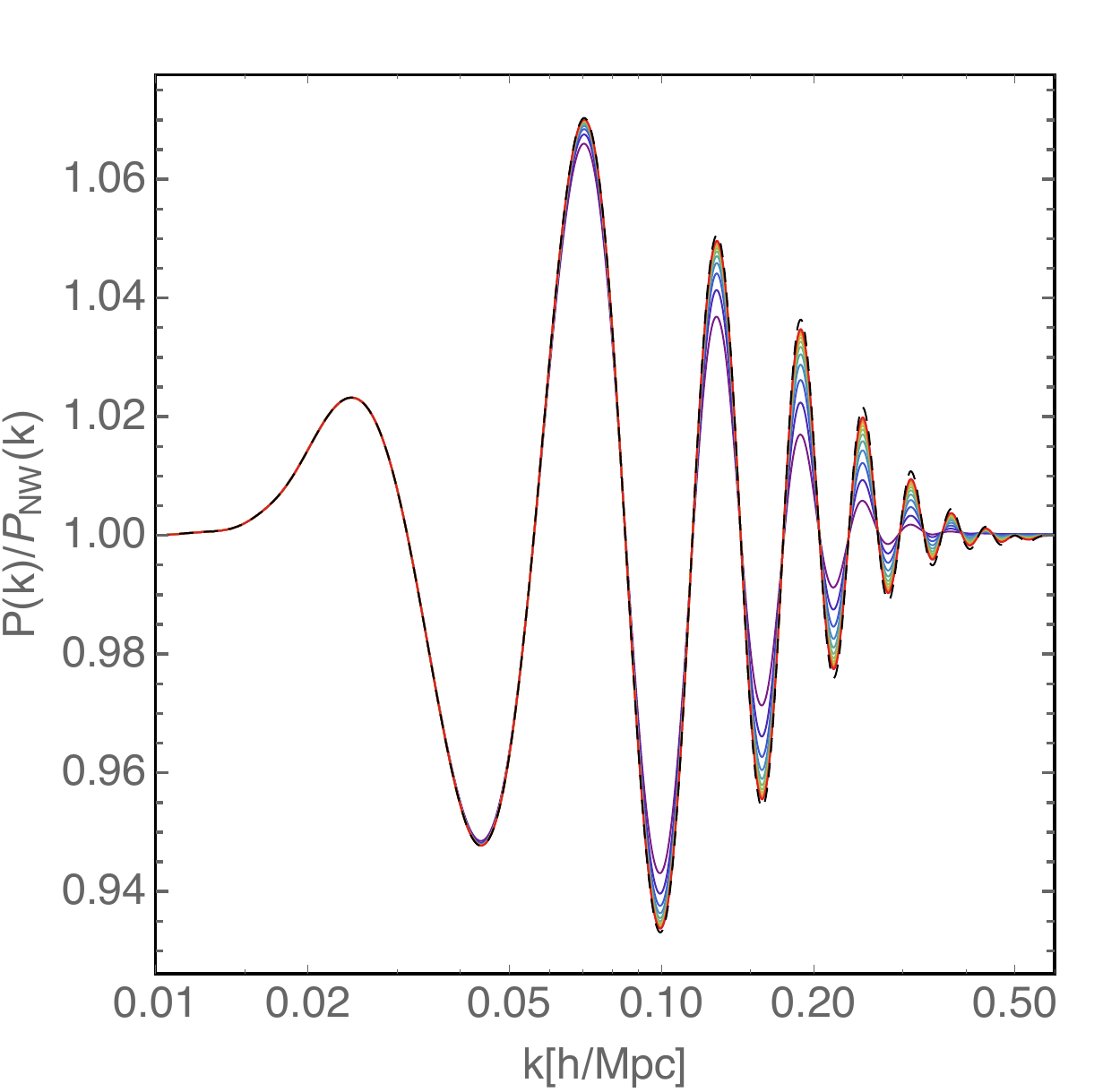}
\caption{We show the ratio between the non-linear power spectrum $ P \left( k \right)$, as defined in Eq.~\eqref{zel}, and the no-wiggle one $P_\text{NW} \left( k \right)$, as defined in Ref.~\cite{Eisenstein:1997ik}, for different redshifts: from $z=0.5$ (violet) to $z=6$ (red) with step-size of $\Delta z=0.5$. 
The black line shows the BAO wiggles for the linear matter power spectrum.
}
\label{fig:PK_nonlinear}
\end{center}
\end{figure}

In Fig.~\ref{fig:PK_nonlinear}, we show the smoothing effect on BAO wiggles induced by non-linearities at different redshifts. By comparing the amplitude of the effects shown in Fig.~\ref{fig:PK_nonlinear} for non-linear dynamics and in Fig.~\ref{fig:PK_NW} for the deflection angle, we clearly notice that the smoothing effect induced by non-linearities is more important at low redshift. Nevertheless its magnitude decreases quickly at high redshift, where non-linear dynamics is less significant at BAO scale. Contrarily, lensing effect is induced by many deflection angles coherently summed along the line of sight. Therefore the effect is roughly proportional to the comoving distance from the source to the observer. We remark that around $z\sim 3$ or $4$ the comoving distance is an half of the distance to the last scattering surface of CMB. This explains the magnitude of the lensing effect for sources around $z\sim 3$ or $4$, we roughly see that its amplitude is a factor $2$ below the CMB lensing.

\begin{figure}[h!]
\begin{center}
\includegraphics[width=5cm]{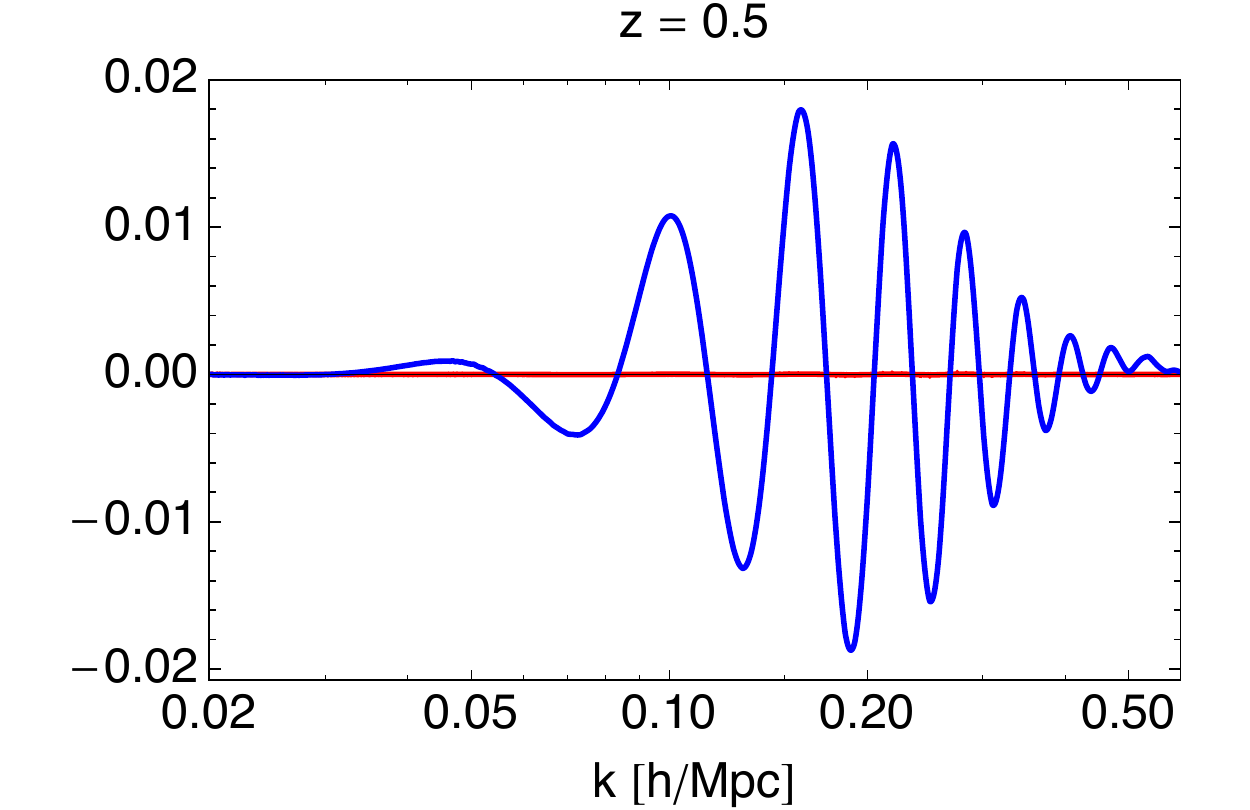}
\includegraphics[width=5cm]{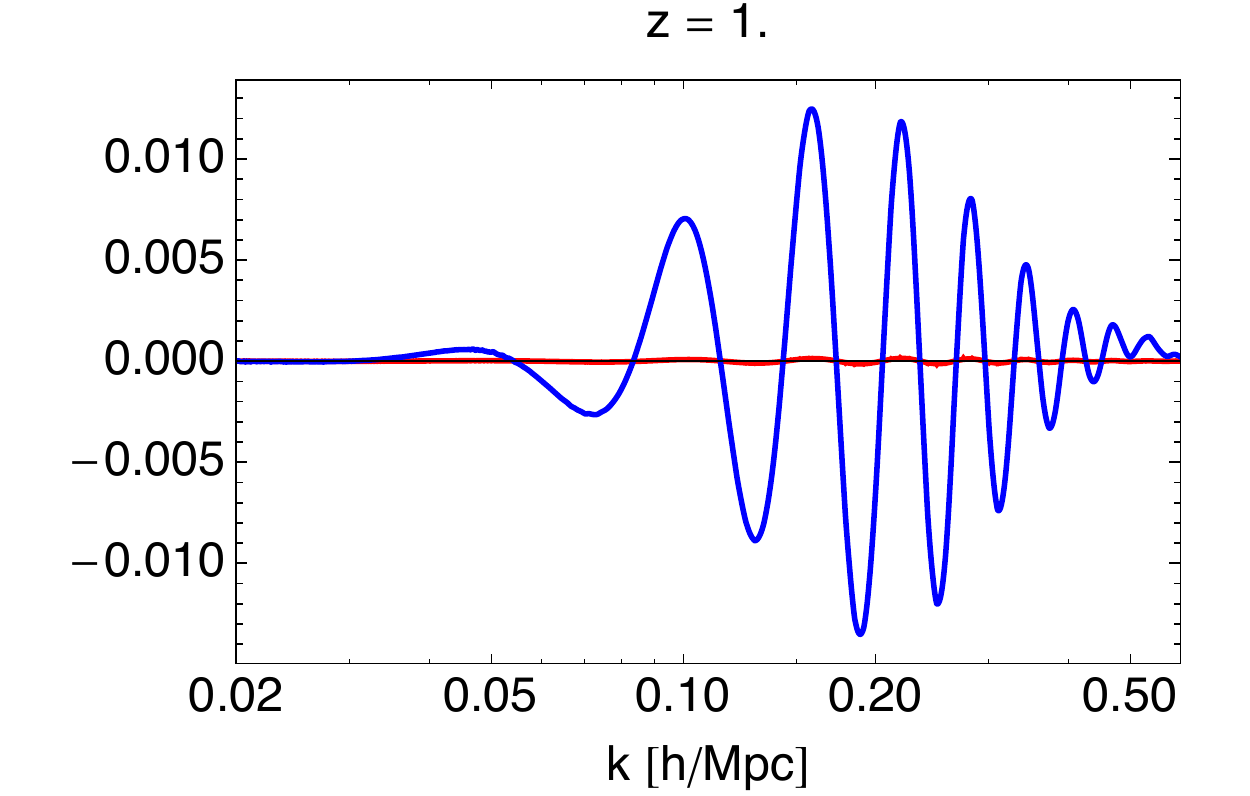}
\includegraphics[width=5cm]{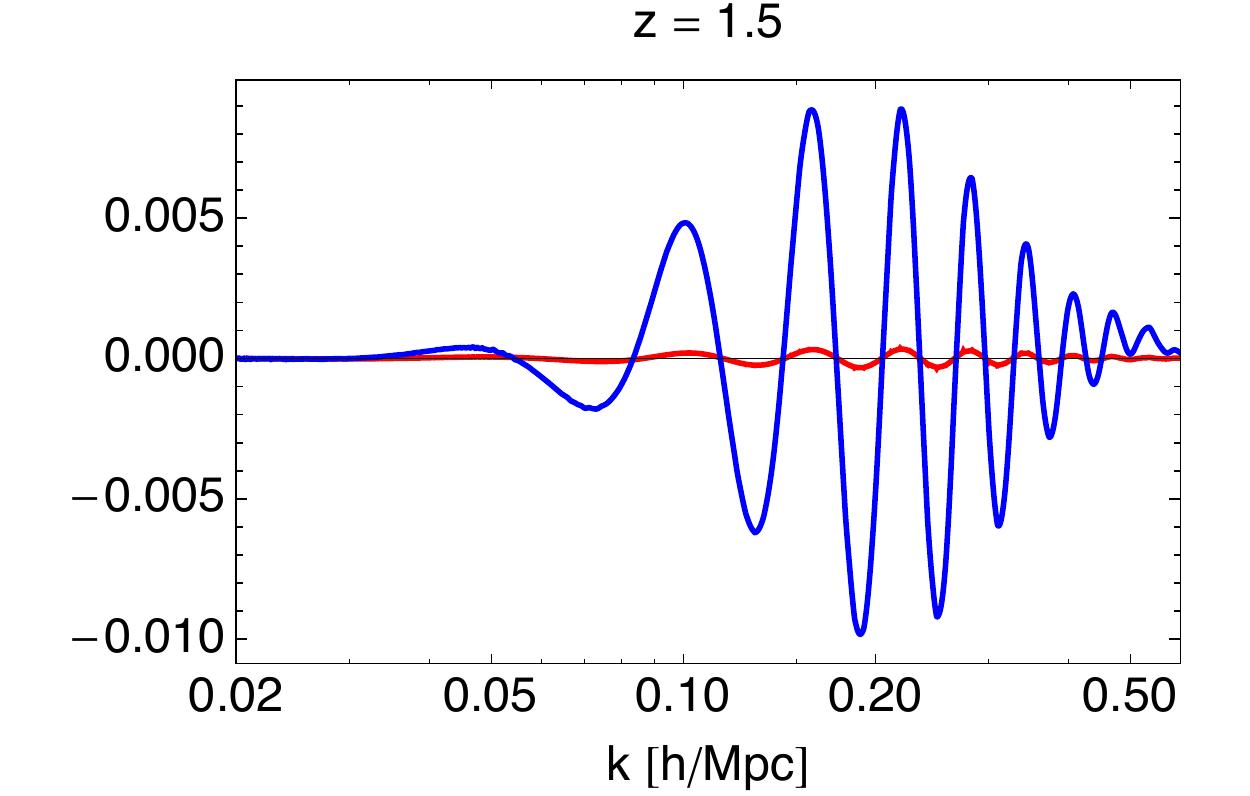}
\\
\includegraphics[width=5cm]{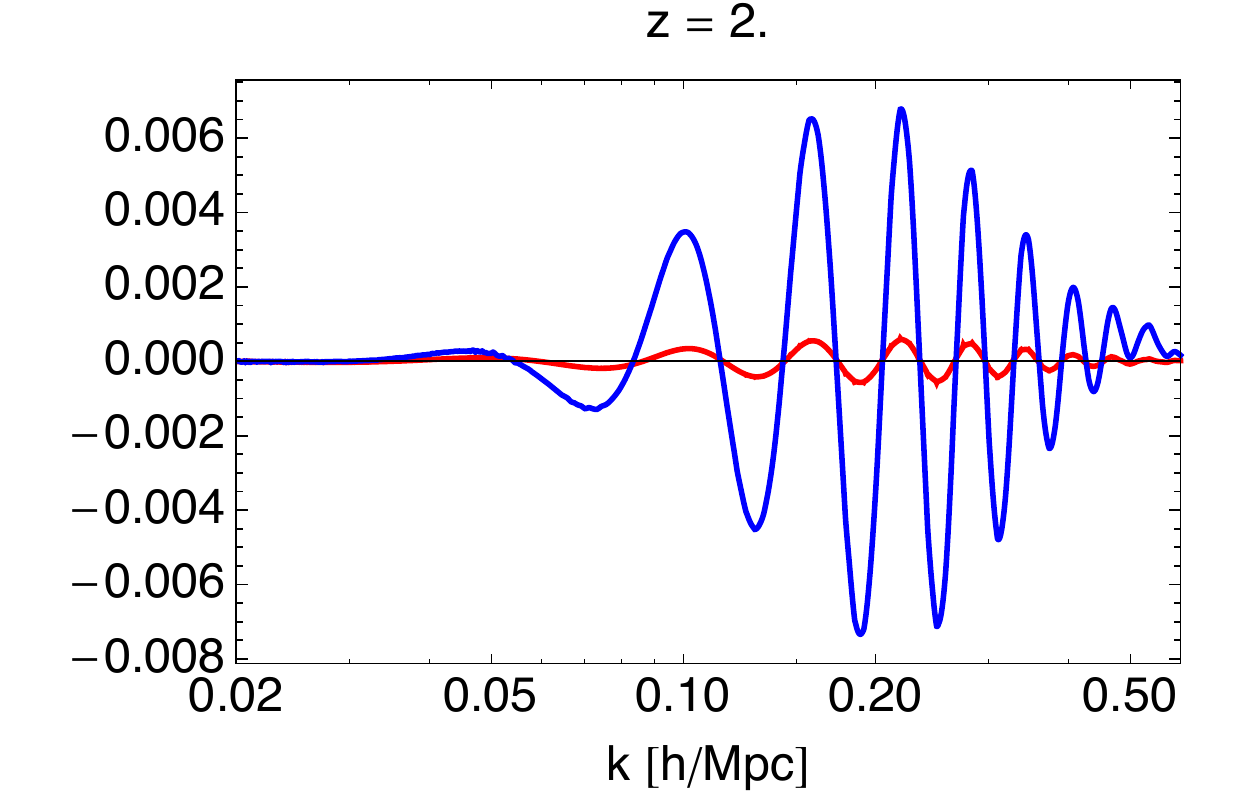}
\includegraphics[width=5cm]{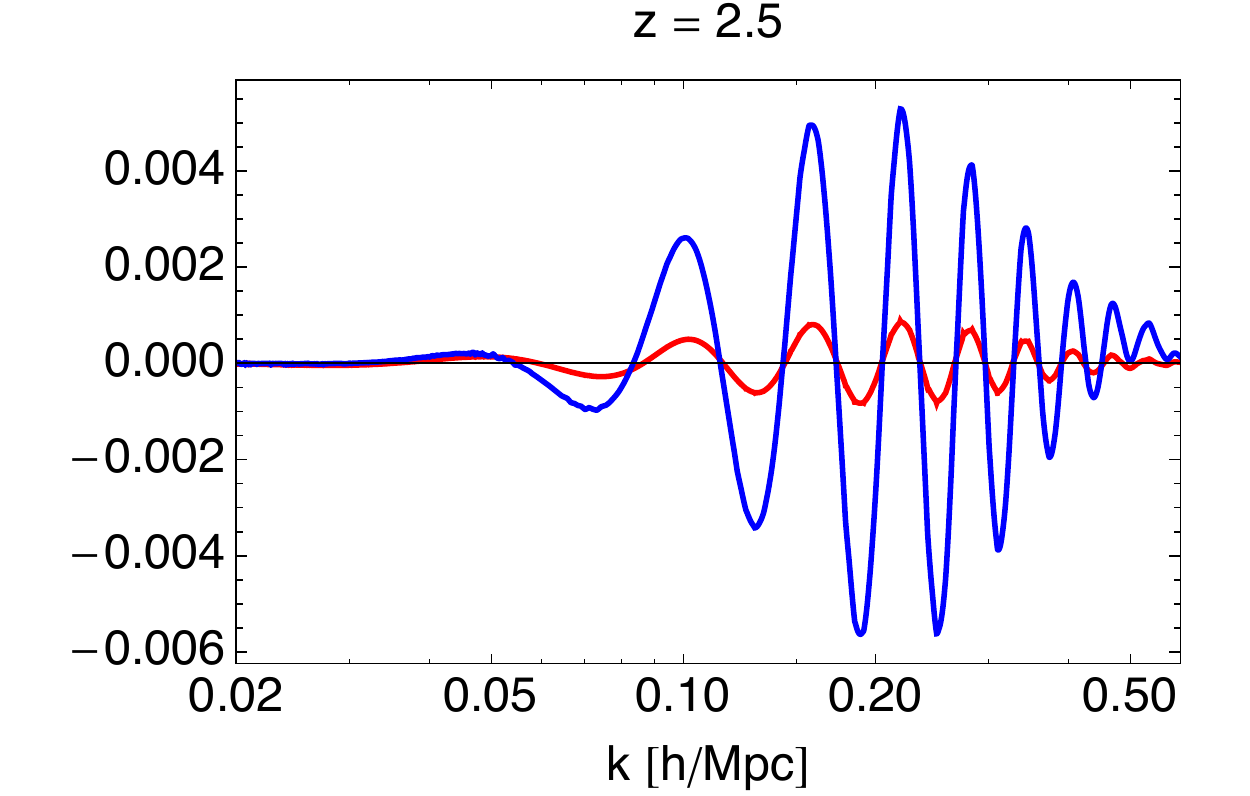}
\includegraphics[width=5cm]{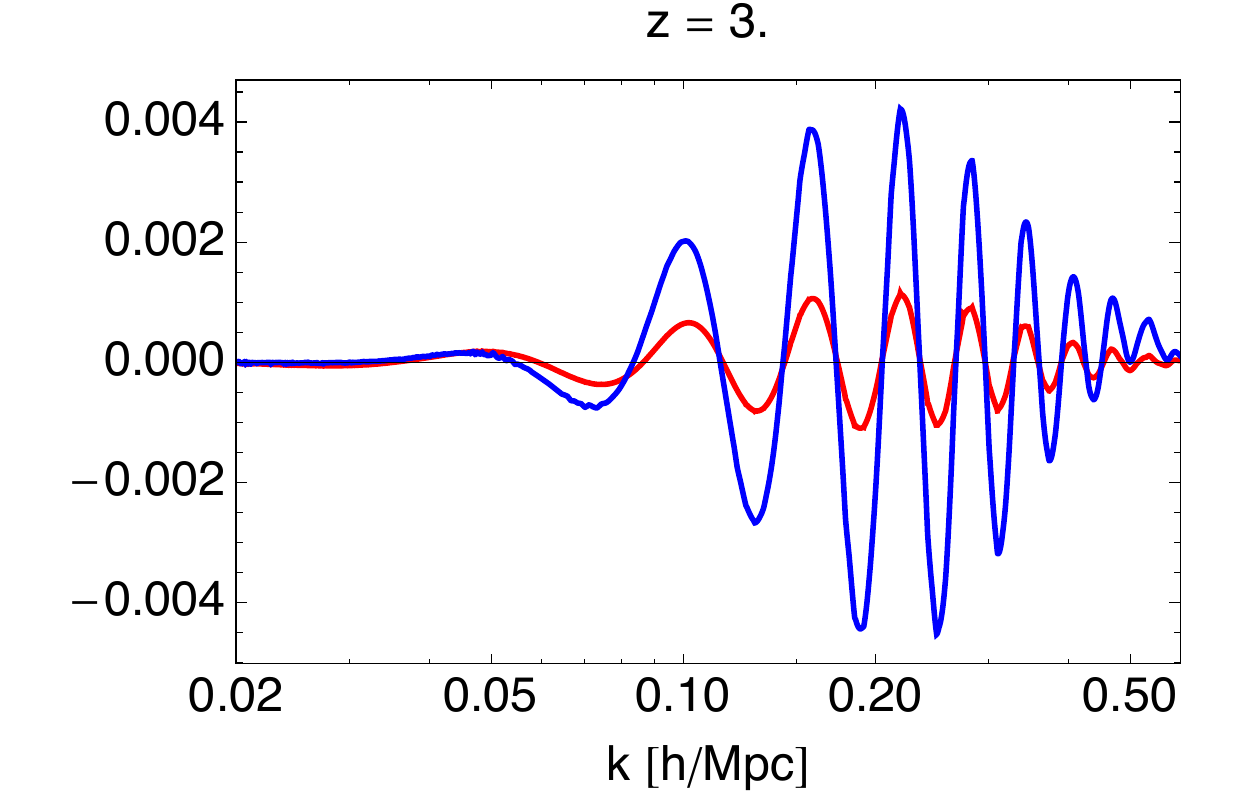}
\\
\includegraphics[width=5cm]{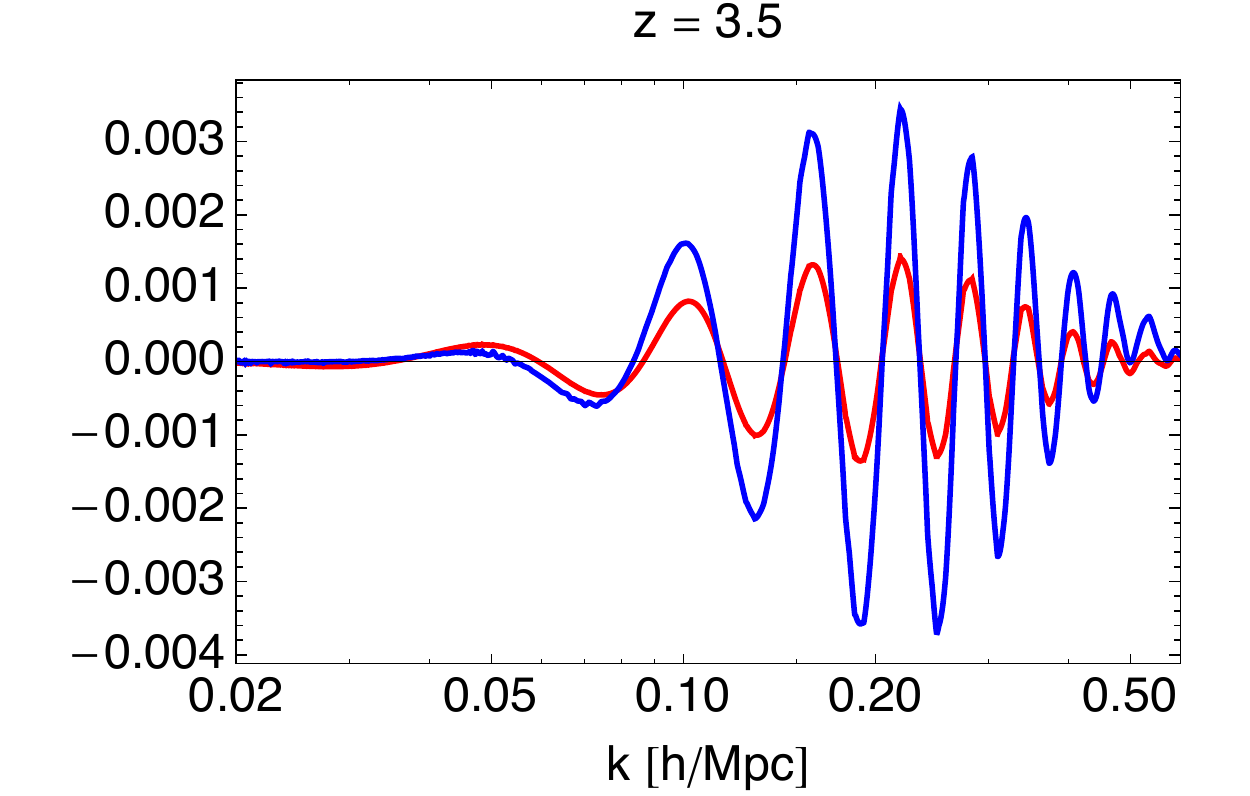}
\includegraphics[width=5cm]{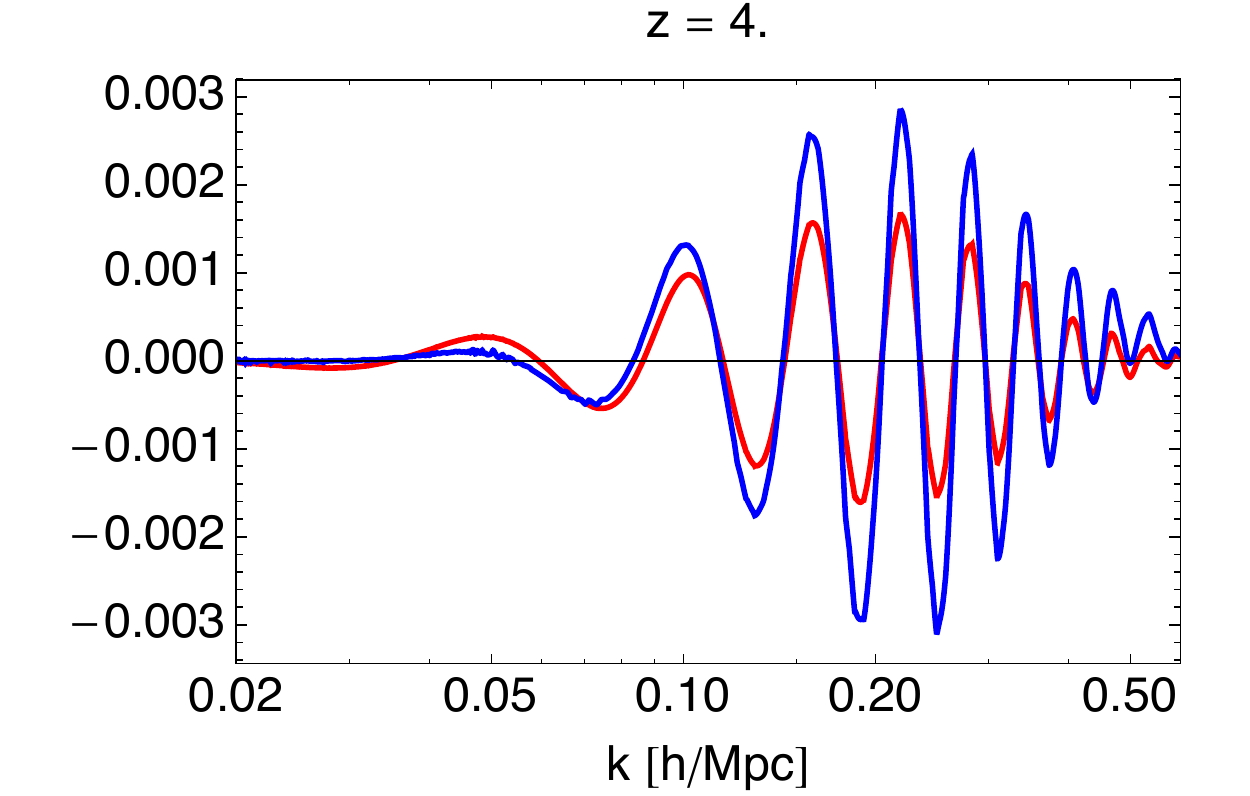}
\includegraphics[width=5cm]{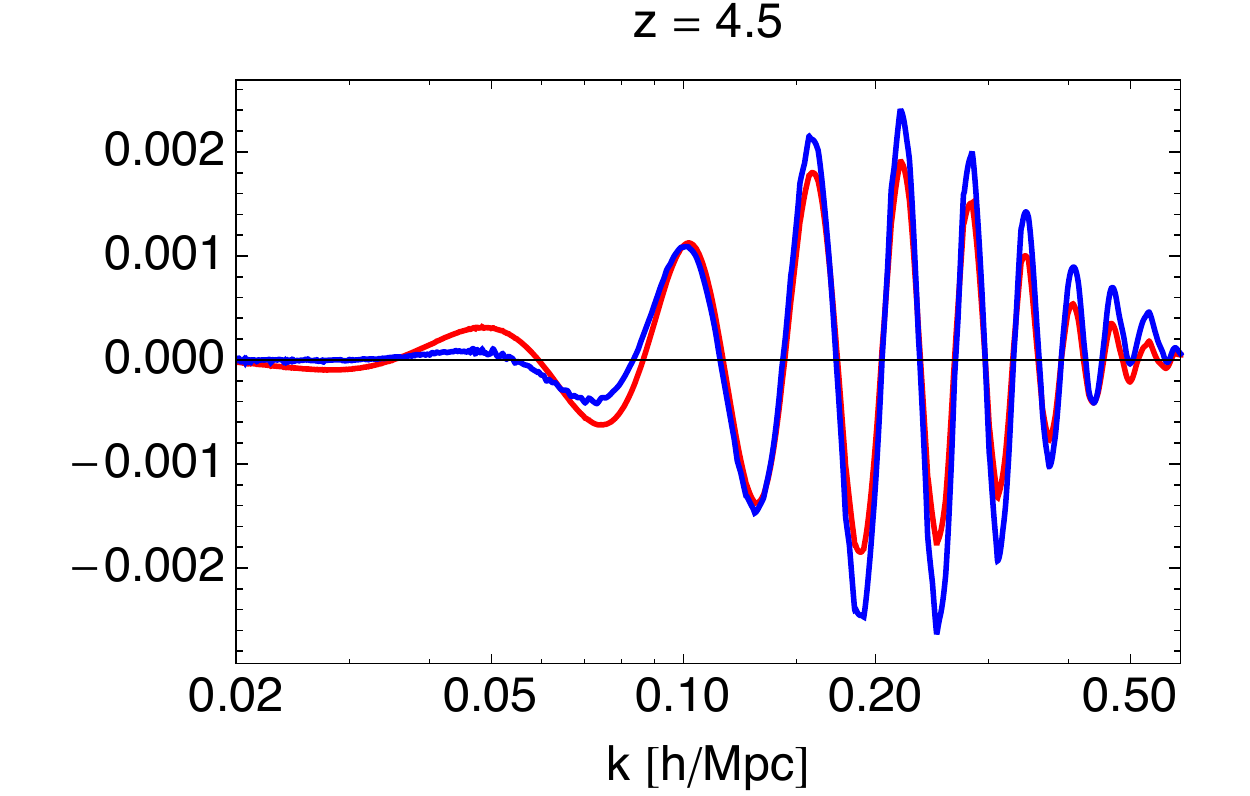}
\\
\includegraphics[width=5cm]{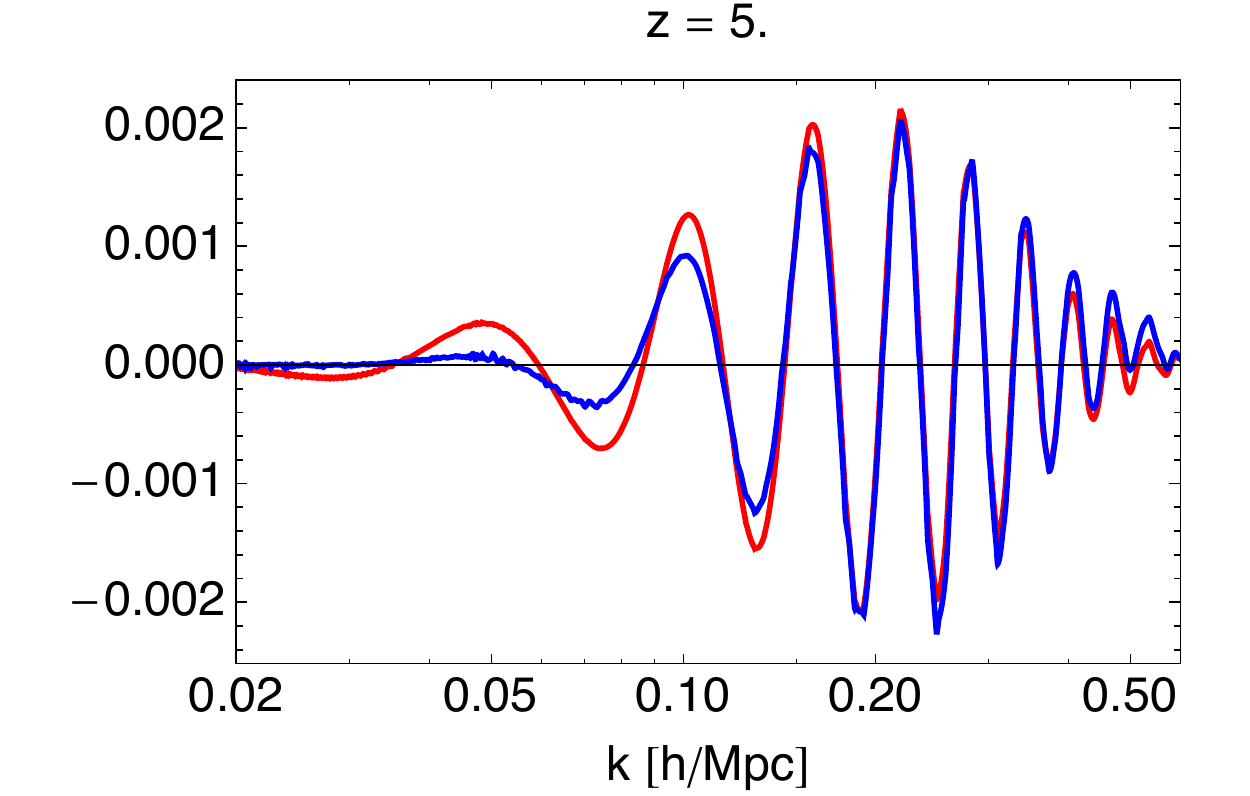}
\includegraphics[width=5cm]{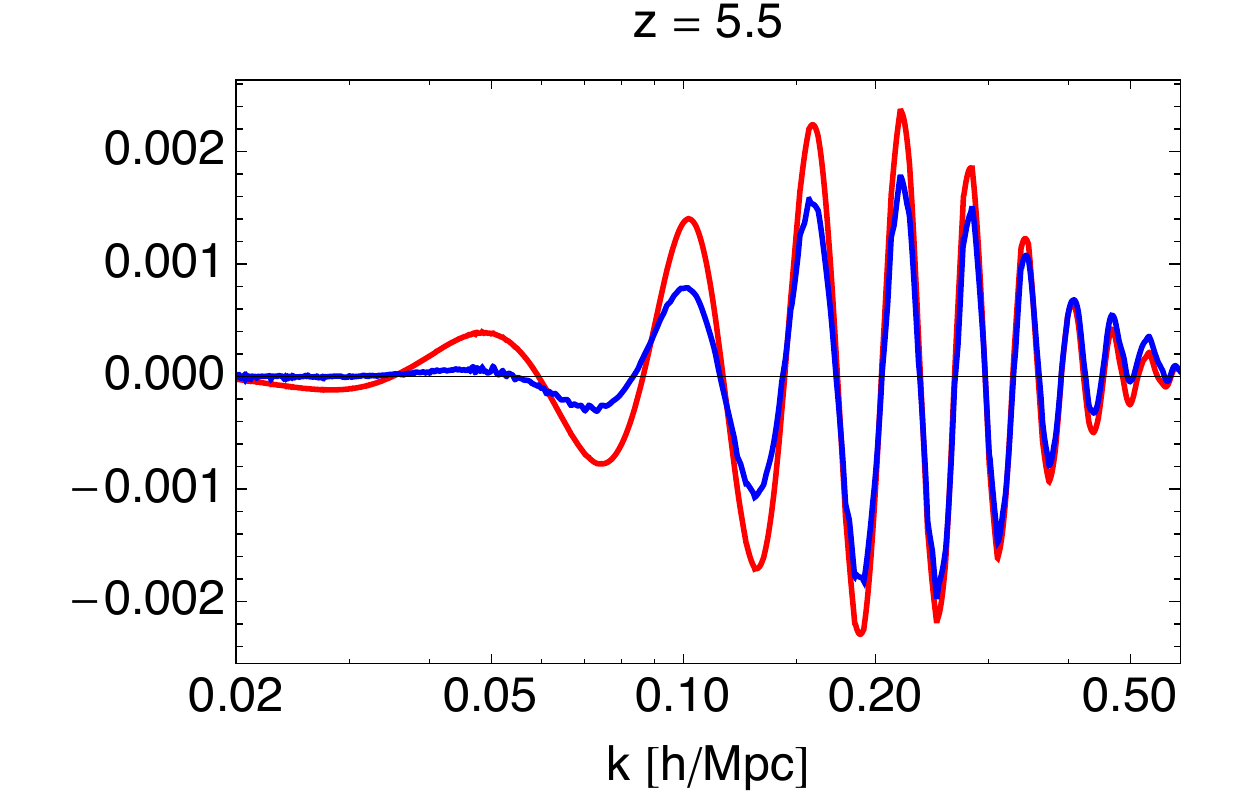}
\includegraphics[width=5cm]{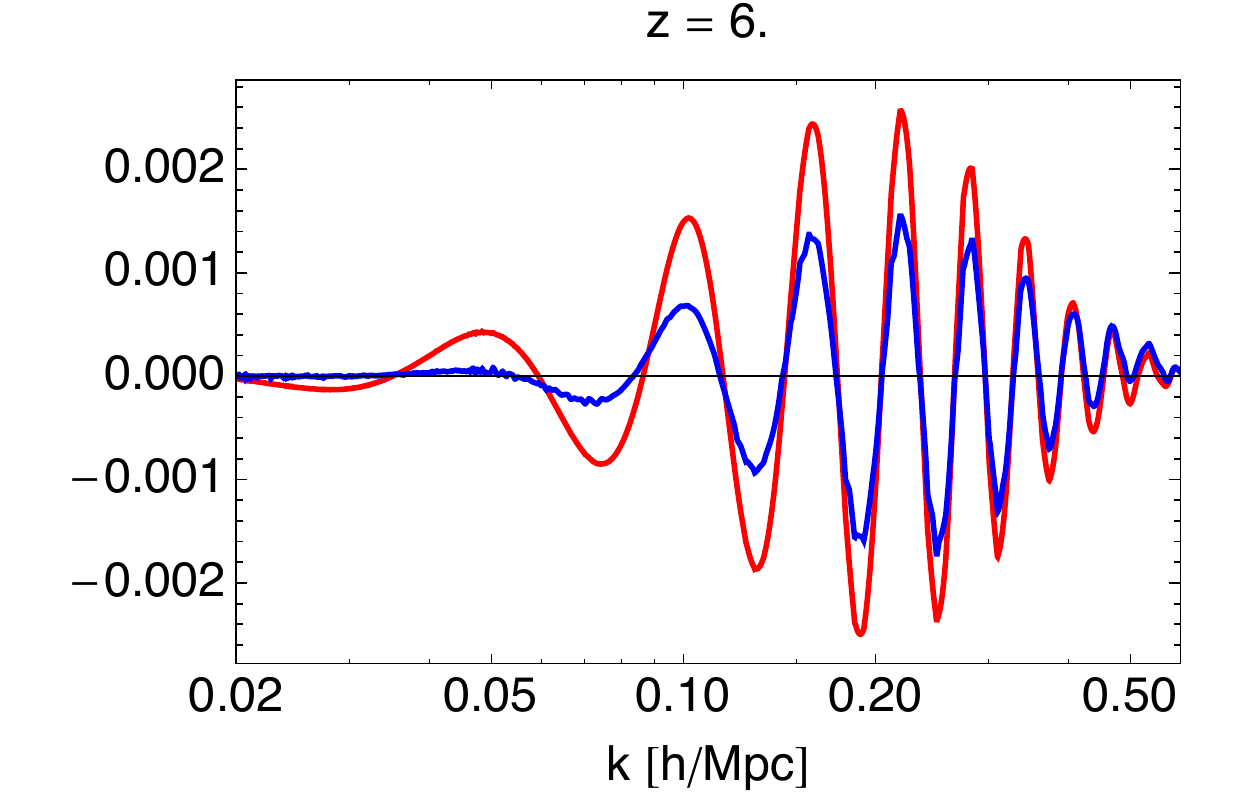}
\caption{We compare the amplitudes of the smoothing of the BAO wiggles induced by lensing potential (red) and by non-linear structure formation (blue) for different redshifts. The magnitude of the effect refers to the amplitude of smooth matter power spectrum.
}
\label{fig:comparison}
\end{center}
\end{figure}

In Fig.~\ref{fig:comparison} we show the magnitude (relative to the smooth matter power spectrum amplitude) of the smoothing effects induced by, respectively, the non-linear and  lensing effects at different redshifts. We plot the relative difference between the full power spectrum with respect to the no-wiggle one. We see the magnitude of the non-linear effects spans from few percent at low redshift to few per-mill at high redshift, while the smoothing due to lensing effect is completely negligible at low redshift and it becomes of few per-mill at high redshift. In particular we notice that at $z\sim 4$ the two effects become comparable and for larger redshifts the lensing effect dominates over the non-linear growth of structures. 

With this comparison we see that the smoothing induced by lensing effect dominates on regimes where the approximation made in previous sections, namely by neglecting the non-Gaussianity of the lensing potential induced by non-linear structure formation, can be safely applied.

Analogously to BAO reconstruction, see e.g.~\cite{Eisenstein:2006nk,Padmanabhan:2012hf}, by knowing the lensing potential from different cosmological probes, we can de-lens the matter power spectrum. Through a de-lensing process, LSS surveys would not be limited by roughly $1\ \text{Mpc}$ resolution at BAO scales for sources at high redshift, as shown in Fig.~\ref{fig:sigma}.
We also remark that BAO reconstruction, i.e. by undoing the non-linear evolution through Zeldovich approximation, does not account for the lensing effect we have computed. Therefore, even if the lensing effect is smaller than the non-linearity at low redshift, it is not compensated by standard data analysis.

%%%%%%%%%%%%%%%%%%%%%%%%%%%%%%%%%%%%%
\section{Conclusions}
\label{sec:conclusions}
In this work we have studied the smoothing effect on BAO wiggles induced on the matter power spectrum in real space by the lensing potential. The smoothing effect has been computed non-perturbatively, under the assumption that the lensing potential obeys Gaussian statistics. We also have estimated the corrections induced by the non-Gaussian nature of the gravitational potential, showing that it induces a very small effect at the relevant redshifts.

We found that lensing introduces a smearing effect with a dispersion of about $1 \ \text{Mpc}$ at BAO scale for sources at redshift $z\sim 2$ or larger. This corresponds to roughly a $1\%$ effect at BAO scale and it is approximately a factor 2 below the lensing effect on the CMB perturbations.
It also introduces a minimal resolution, which can be improved only by knowing the lensing potential and by de-lensing the matter power spectrum. By comparing with a no-wiggle power spectrum we have shown the effect on the BAO wiggles. In particular, the lensing potential reduces the amplitude of BAO wiggles, by affecting the matter power spectrum of about $0.1 \%$ at $z\sim 2$ and $0.2 \%$ at $z\sim 4$. For sources at larger redshifts the suppression is even larger.
We remind that our results apply also for 21cm matter power spectrum.

We have finally compared the smoothing effects on BAO wiggles induced by lensing and by non-linear structure formation. While at low redshift the non-linear dynamics of gravity leads to the dominant smearing effect on the BAO wiggles, the two effects become comparable at $z\sim 4$ and at larger redshifts the lensing effect dominates over the non-linearities.
So, if on one hand, sources at high redshift are less affected by non-linearities, on the other hand, they are more sensitive to the smearing effect due to the lensing potential.

In the literature lensing effects on LSS observables have been studied in terms of the cosmic magnification. Contrarily to the deflection angle, which we have studied in this work, cosmic magnification enters at first order in perturbation theory. Different works have quantified the amplitude of this effect and stressed its importance for future surveys. Nevertheless, its amplitude depends on the galaxy and magnification biases. Hence its effect can be degenerated with these terms. Whereas, the deflection angle is a purely geometrical effect and it is not degenerate with any bias factors, but it is sensitive only to the metric perturbations through the lensing potential.

%%%%%%%%%%%%%
\section*{Acknowledgments}
I am grateful to Uro\v s Seljak for the useful discussions and for suggesting the topic.
I would like to thank Ruth Durrer and Francesco Montanari for useful comments on the manuscript.
ED is supported by the ERC Starting Grant cosmoIGM and by INFN/PD51 INDARK grant.

%%%%%%%%%%%%%%%%%%%%%%%%%%%%%%%%%%%
\appendix
\section{Non-Gaussian LSS correction}
\label{app:LSS_NG}
In the derivation of the lensed matter power spectrum we have assumed that the deflection angle obeys Gaussian statistics, through Eq.~\eqref{Gaussian_trick}. In this appendix we are interested in estimating the correction induced by the non-Gaussianity of the non-linear structure formation at low redshift. A similar calculation has been recently performed in the CMB framework, see e.g.~\cite{Pratten:2016dsm,Lewis:2016tuj,Marozzi:2016qxl,Marozzi:2016und}, showing very small corrections. Here we repeat the same calculation for the lensed matter power spectrum, defined as
\be
\langle \tilde \delta \left( \bk \right)  \tilde \delta \left( \bk' \right) \rangle = \delta_D^{(3)} \left( \bk + \bk' \right)  \left( \tilde P_\text{Gaussian} \left( k \right) + \Delta P \left( k \right) \right),
\ee
where $\tilde P_\text{Gaussian} \left( k \right) $ coincides with the lensed power spectrum derived in Eq.~\eqref{lensed_PK}.

The non-Gaussian contributions can not be resummed and therefore we need to expand perturbatively the lensed density fluctuation
\bea
&& \tilde \delta \left( \bx \right) = \delta \left( \bx + \delta \bx \right) 
\nonumber \\
&&\sim 
\delta \left( \bx \right) + \delta x^i \nabla_i \delta \left( \bx \right) 
+\frac{1}{2} \delta x^i \delta x^j \nabla_i \nabla_j \delta  \left( \bx \right)+ \frac{1}{6} \delta x^i \delta x^j \delta x^k \nabla_i \nabla_j \nabla_k \delta  \left( \bx \right)+ \mathcal{O} \left( \left( \delta \bx \right)^4 \right) \, . \qquad
\eea
We can now compute the lensed correlation function
\bea
&& \hspace{-1cm}
\tilde \xi \left( \Delta x_\perp, \Delta x_\parallel \right)= \langle \tilde  \delta \left( \bx_1  \right) \tilde \delta \left( \bx_2 \right) \rangle = \langle \delta \left( \bx_1 + \delta \bx_1 \right) \delta \left( \bx_2 + \delta \bx_2 \right) \rangle
\nonumber \\
&\simeq& \langle \delta \left( \bx_1 \right) \delta \left( \bx_2 \right) \rangle 
\nonumber \\
&&
+ \frac{1}{2} \left( \langle \delta x_1^i \delta x_1^j \rangle \langle \nabla_i \nabla_j \delta \left( \bx_1\right) \delta \left( \bx_2 \right) \rangle 
+ \langle \delta x_2^i \delta x_2^j \rangle \langle \nabla_i \nabla_j \delta \left( \bx_2\right) \delta \left( \bx_1 \right) \rangle 
\right.
\nonumber \\
&&
\left. \qquad
+2  \langle \delta x_1 ^i \delta x_2^j  \rangle \langle \nabla_i \delta \left( \bx_1 \right) \nabla_j \delta \left( \bx_2 \right) \rangle
\right)
\nonumber \\
&&
+ \frac{1}{2} \left( 
\langle \delta x_1^i \delta x_1^j \delta x_2^k \rangle \langle \nabla_i \nabla_j \delta \left( \bx_1 \right) \nabla_k \delta \left( \bx_2 \right) \rangle
+
\langle \delta x_2^i \delta x_2^j \delta x_1^k \rangle \langle \nabla_i \nabla_j \delta \left( \bx_2 \right) \nabla_k \delta \left( \bx_1 \right) \rangle
\right) \,  ,\
\label{lens_corr_perturbative}
\eea
where we have always assumed that the sources and the lensing potential are uncorrelated.
The first terms in this perturbative expansion which are not captured by the exponential resummation are described by the last line of Eq.~\eqref{lens_corr_perturbative}. These terms induce a non-vanishing bispectrum which determines the LSS non-Gaussian correction $\Delta P \left( k \right) $ to the resummed lensed matter spectrum.

We compute now the 3-point function
\bea
&&\langle \delta x^i_1 \delta x_2^j \delta x_3^k \rangle =
 - 8 i \int_0^{r_1} dr' \left( r' -r_1\right)  \int_0^{r_2} dr'' \left( r'' -r_2 \right)  \int_0^{r_3} dr''' \left( r''' -r_3 \right)
\nonumber \\
&&
  \int  \frac{d^3 k_1 d^3 k_2 d^3 k_3}{(2 \pi)^{9/2}} \langle \Phi_W \left( \bk_1 \right) \Phi_W \left( \bk_2 \right) \Phi_W \left( \bk_3 \right)  \rangle k_{1\perp}^i k_{2\perp}^j k_{3\perp}^k e^{i \left( \bk_1 \cdot \bx'_1 +\bk_2 \cdot \bx'_2  +\bk_3 \cdot \bx'_3\right)}
  \nonumber \\
  &=&
   - 8 i \int_0^{r_1} dr' \left( r' -r_1\right)  \int_0^{r_2} dr'' \left( r'' -r_2 \right)  \int_0^{r_3} dr''' \left( r''' -r_3 \right)
\nonumber \\
&&
  \int  \frac{d^3 k_1 d^3 k_2 d^3 k_3}{(2 \pi)^{6}} B^{\phi \phi \phi} \left( \bk_1 , \bk_2 , \bk_3 \right) \delta_D^{(3)} \left( \bk_1 + \bk_2 + \bk_3 \right)
   k_{1\perp}^i k_{2\perp}^j k_{3\perp}^k e^{i \left( \bk_1 \cdot \bx'_1 +\bk_2 \cdot \bx'_2  +\bk_3 \cdot \bx'_3\right)}
   \nonumber \\
   &\sim&
    8 i \int_0^{\text{min}(r_1, r_2 ,r_3)} dr'   \left( r' -r_1\right)  \left(r' -r_2 \right)  \left( r' -r_3 \right)
        \int  \frac{d^2 k_{1 \perp} d^2 k_{\perp 2}}{\left( 2 \pi \right)^4}
   \nonumber \\
   &&
 B^{\phi \phi \phi} \left( \bk_{\perp 1}, \bk_{\perp 2} , - \bk_{\perp 1} - \bk_{\perp 2} \right)  
       k_{1\perp}^i k_{2\perp}^j \left( k_{1\perp}^k  + k_{2\perp}^k  \right) 
       e^{i \bk_{1 \perp} \cdot \left( \bx'_1 - \bx'_3 \right) } e^{i \bk_{2 \perp} \cdot \left( \bx'_2 - \bx'_3 \right) } \, ,
       \eea
where we have introduced the Weyl potential $\Phi_W = \left( \psi + \phi \right) /2$ and its bispectrum $B^{\phi \phi \phi}$. We have also applied Limber approximation.

Then, under the approximation $r'- r_1  \sim  r'- r_2$, we have
\bea
\Delta P(k) &=& 8 
\int d^2 \Delta x_\perp d \Delta x_\parallel e^{-i \bk_\perp \cdot {\bxx_\perp}}
 \int_0^{r=\text{min}(r_1, r_2 )} dr' \left(r' -r  \right)^3
        \int  \frac{d^2 k_{ \perp 1} d^2 k_{\perp 2}}{\left( 2 \pi \right)^4}
   \nonumber \\
   &&
 B^{\phi \phi \phi} \left( \bk_{\perp 1}, \bk_{\perp 2} , - \bk_{\perp 1} - \bk_{\perp 2} \right)  
       k_{1\perp}^i k_{2\perp}^j \left( k_{1\perp}^k  + k_{2\perp}^k  \right) 
       e^{- i \left( \bk_{1 \perp}+\bk_{2 \perp} \right) \cdot \bxx_\perp \frac{r'}{r }} 
       \nonumber \\
       &&
       \int \frac{d^3 k_3 d^3k_4}{\left( 2 \pi \right)^3} k_3^i k_3^j k^k_4 e^{i \left(  \bk_3 \cdot \bx_1 + \bk_4 \cdot \bx_2 \right)} P\left( k_3 \right) \delta_D^{(3)} \left( \bk_3 + \bk_4 \right)
       \nonumber \\
       &=& -8 
\int d^2 \Delta x_\perp d \Delta x_\parallel e^{-i \bk_\perp \cdot {\bxx_\perp}}
 \int_0^{r=\text{min}(r_1, r_2 )} dr'  \left(r' -r  \right)^3
        \int  \frac{d^2 k_{ \perp 1} d^2 k_{\perp 2}}{\left( 2 \pi \right)^4}
   \nonumber \\
   &&
 B^{\phi \phi \phi} \left( \bk_{\perp 1}, \bk_{\perp 2} , - \bk_{\perp 1} - \bk_{\perp 2} \right)  
       k_{1\perp}^i k_{2\perp}^j \left( k_{1\perp}^k  + k_{2\perp}^k  \right) 
       e^{- i \left( \bk_{1 \perp}+\bk_{2 \perp} \right) \cdot \bxx_\perp \frac{r'}{r }}        \nonumber \\
       &&
       \int \frac{d^3 k_3 }{\left( 2 \pi \right)^3} k_3^i k_3^j k^k_3 e^{- i   \bk_3 \cdot  \bxx} P\left( k_3 \right) 
 \nonumber \\
 &=&
 -8 
\int d^2 \Delta x_\perp  e^{-i \bk_\perp \cdot {\bxx_\perp}}
 \int_0^{r=\text{min}(r_1, r_2 )} dr'  \left(r' -r  \right)^3
        \int  \frac{d^2 k_{ \perp 1} d^2 k_{\perp 2} d^2 k_{\perp 3}}{\left( 2 \pi \right)^6}
   \nonumber \\
   &&
 B^{\phi \phi \phi} \left( \bk_{\perp 1}, \bk_{\perp 2} , - \bk_{\perp 1} - \bk_{\perp 2} \right)  
            e^{- i \left( \bk_{1 \perp}+\bk_{2 \perp} \right) \cdot \bxx_\perp \frac{r'}{r }}    e^{- i   \bk_{\perp 3} \cdot  \bxx_\perp} P\left( k^{\perp }_3 \right)   \nonumber \\
       &&  \bk_{\perp 1}\cdot \bk_{\perp 3}    \bk_{\perp 2}\cdot \bk_{\perp 3}    \left( \bk_{\perp 1}+ \bk_{\perp 2} \right)\cdot \bk_{\perp 3}   
   \nonumber \\
   &=&
    -8 
 \int_0^{r=\text{min}(r_1, r_2 )} dr' \left( \frac{r}{r'}\right)^2  \left(r' -r  \right)^3
        \int  \frac{d^2 k_{ \perp 1} d^2 k_{\perp 3}}{\left( 2 \pi \right)^4}
   \nonumber \\
   &&
 B^{\phi \phi \phi} \left( \bk_{\perp 1}, -\frac{r}{r'}\left( \bk_\perp + \bk_{\perp 3} \right) - \bk_{\perp 1} , \frac{r}{r'}\left( \bk_\perp + \bk_{\perp 3} \right)  \right)  
         P\left( k^{\perp }_3 \right)   \nonumber \\
       && \left(  \bk_{\perp 1}\cdot \bk_{\perp 3} \right)  
       \left( \frac{r}{r'}\left( \bk_\perp +  \bk_{\perp 3} \right) + \bk_{\perp 1} \right) \cdot \bk_{\perp 3}
       \frac{r}{r'}  \left( \bk_{\perp }+ \bk_{\perp 3} \right)\cdot \bk_{\perp 3}       
         \nonumber \\
   &=&
    -8 
 \int_0^{r=\text{min}(r_1, r_2 )} dr'   \left( r' -r  \right)^3 \left( \frac{r}{r'} \right)^7
        \int  \frac{d^2 k_{ \perp 1} d^2 k_{\perp 3}}{\left( 2 \pi \right)^4}
   \nonumber \\
   &&
 B^{\phi \phi \phi} \left( \frac{r}{r'} \bk_{\perp 1}, -\frac{r}{r'}\left( \bk_\perp + \bk_{\perp 3}  + \bk_{\perp 1} \right) , \frac{r}{r'}\left( \bk_\perp + \bk_{\perp 3} \right)  \right)  
         P\left( k^{\perp }_3 \right)   \nonumber \\
       && \left(  \bk_{\perp 1}\cdot \bk_{\perp 3} \right)  
       \left( \left( \bk_\perp +  \bk_{\perp 3} \right) + \bk_{\perp 1} \right) \cdot \bk_{\perp 3}
  \left( \bk_{\perp }+ \bk_{\perp 3} \right)\cdot \bk_{\perp 3}       \, .
  \label{DeltaPK}
\eea

\begin{figure}[htb!]
\begin{center}
\includegraphics[width=7cm]{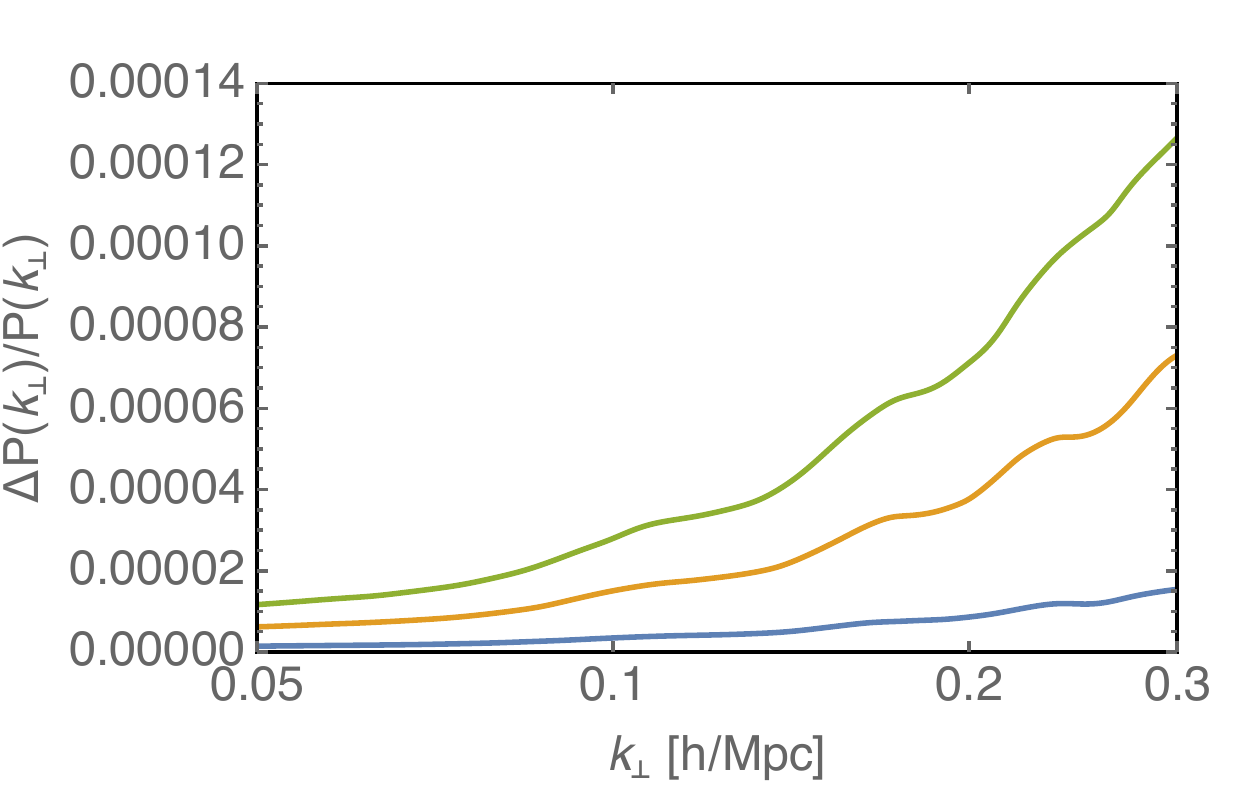}
\caption{We show the relative corrections induced by the non-Gaussian nature of the gravitational potential at different redshifts: $z=2$ (blue), $z=4$ (orange) and $z=6$ (green).
}
\label{fig:NG}
\end{center}
\end{figure}

We integrate numerically Eq.~\eqref{DeltaPK} with the Suave\footnote{\url{http://www.feynarts.de/cuba/}} Monte Carlo integrator.
In Fig.~\ref{fig:NG} we show the relative corrections induced by the non-Gaussian nature of the gravitational potential at different redshifts. We notice that the amplitude grows with redshift. Indeed, even if the non-Gaussian nature of LSS is more important at low redshift, the full effect is integrated along the line of sight. The corrections are always much smaller than the per-mill level, i.e.~the largest correction induced by the full Gaussian part. This shows that for sources at high redshift, where the lensing effect is more relevant, the non-Gaussian corrections are negligible.

It is interesting to notice that the corrections increase at small scales. This can be due to the break down of the perturbative expansion. Indeed we do expect the perturbative expansion to converge slower than CMB, see e.g.~\cite{Mandel:2005xh}. Therefore a more accurate estimation of the effects induced by non-Gaussianity would require a light-tracing process through N-body simulations.

%%%%%%%%%%%%%%%%%%%%%%%%%%%%%%%%%%%
\section{Lensed angular power spectrum}
\label{app:a}

In Section~\ref{sec:PK} we have derived the lensed matter power spectrum under some approximations. Namely, we have considered that the correlation induced by longitudinal modes $\delta x_\parallel$ are subdominant compare the transversal modes $\delta \bx_\perp$ and we have adopted Limber approximation. Here we show that we can derive the analogous result in terms of the redshift dependent angular power spectra $C_\ell$. Even if not commonly used, apart to consistently include all the relativistic corrections for galaxy number counts, see e.g.~\cite{Challinor:2011bk,Bonvin:2011bg} for the power spectrum and~\cite{DiDio:2014lka,DiDio:2015bua} for the bispectrum, the harmonic space offers the advantage to describe the statistical properties of the perturbations in terms of observable quantities only. This means that we can generalize the results previously derived to include other effects, like redshift-space distortions, Doppler or potential terms.
In a relativistic framework the galaxy number counts~\cite{Challinor:2011bk,Bonvin:2011bg,DiDio:2013bqa} is described by
\bea \label{number_counts}
 \Delta \left( \bn, z \right) &=& b \delta^\text{sync} +(5s -2)\Phi + \Psi + \frac{1}{\HH}
\left[\dot \Phi+\dd_r(\ndv)\right] +  \left( f_\text{evo} - 3 \right) \HH v\nonumber \\  &&
+ \left(\frac{{\dot\HH}}{\HH^2}+\frac{2-5s}{r_S\HH} +5s-f_{\rm evo}\right)\left(\Psi+\ndv+ 
 \int_0^{r_S}\hspace{-0.3mm}dr(\dot \Phi+\dot \Psi)\right) 
   \nonumber \\  &&  \label{DezNF}
+\frac{2-5s}{2r_S}\int_0^{r_S}\hspace{-0.3mm}dr \left[2-\frac{r_S-r}{r}\Delta_\Om\right] (\Phi+\Psi) \, ,
\eea
where we have introduced the magnification bias
\be \label{magnification_bias}
s= - \left. \frac{2}{5} \frac{\partial \ln \bar n \left( z, \ln L \right)}{\partial \ln L} \right|_{\bar L}
\ee
where $\bar L$ denotes the threshold luminosity of the survey and $\bar
n$ is the background number density, and the evolution bias
\be \label{fevo}
f_\text{evo} = \frac{\partial \ln \left( a^3 \bar n \right)}{\HH \partial t}
= 3 - \left( 1+ z \right) \frac{d \ln \bar n  } {dz} \, .
\ee

Then, by working on flat-sky approximation\footnote{Let us remark that we have used implicitly flat-sky approximation in the derivation in Section~\ref{sec:PK}. This approximation works better at high redshift, because the BAO scale translates into a smaller angular scale than at low redshift. Hence it should work well at redshifts where the corrections induced by deflection angle are more relevant.} we have
\bea
\Delta \left( \bx, z \right) &=& \frac{1}{2 \pi} \int d^2 \ell \ \Delta \left( \bell, z \right) e^{i \bx \cdot \bell}\, , \\
\Delta \left( \bell, z \right) &=& \frac{1}{2 \pi} \int d^2 x \ \Delta \left( \bx, z \right) e^{-i \bx \cdot \bell}\, ,
\eea
where $\bx$ and $\bell$ are 2-dimensional vectors on the space orthogonal to the line of sight $\bn$. Hence $\bx$ and $\bell$ are dimensionless. Nevertheless we use the same notation we have previously used, and position $\bx$ adopted in this Appendix is trivially related through a factor $r\left( z\right)$ (i.e.~the comoving distance to redshift $z$) to the one defined in the main text. Similarly to Eq.~\eqref{corr_function2} we compute the lensed correlation function
\bea
\tilde \xi \left( z_1 ,z_2 , \Delta x_\perp \right) &=&  \langle \tilde \Delta \left( \bx_1, z_1 \right) \tilde \Delta \left( \bx_2 , z_2 \right) \rangle = \langle  \Delta \left( \bx_1+ \delta \bx_{\perp 1}, z_1\right)  \Delta \left( \bx_2 + \delta \bx_{\perp 2} , z_2 \right) \rangle
\nonumber \\
&=&
 \frac{1}{\left( 2 \pi \right)^2} \int d^2 \ell \  e^{- i \bell \cdot \bxx_\perp}   e^{- \langle \left[ \bell \cdot  \left( \delta \bx_{\perp 1}-   \delta \bx_{\perp 2}\right) \right]^2 \rangle/2} C^{\Delta \Delta}_\ell \left( z_1 , z_2 \right)
 \nonumber \\
 &=&
  \frac{1}{\left( 2 \pi \right)^2} \int d^2 \ell \  e^{- i \bell \cdot \bxx_\perp}   e^{- \langle \left[ \bell \cdot  \left( \nabla_\perp \psi_1 -   \nabla_\perp \psi_2 \right) \right]^2 \rangle/2} C^{\Delta \Delta}_\ell \left( z_1 , z_2 \right)
\eea
where we have assumed statistical angular homogeneity and isotropy by introducing the redshift dependent angular power spectra through
\be
\langle\Delta \left( \bell , z_1 \right) \Delta \left( \bell', z_2 \right) \rangle = \delta^{(2)}_D \left( \bell + \bell' \right) C^{\Delta\Delta}_\ell \left( z_1 , z_2 \right) \, .
\ee
We remark again that we have implicitly assumed that the lensing potential is described by Gaussian statistics and that the deflection angles and the sources are uncorrelated. If $\Delta$ denotes the full number counts, then we have also some non-local terms, e.g. cosmic magnification, ISW and time-delay effects. While ISW and time-delay effects are usually negligible, cosmic magnification may introduce some long range correlation along the line of sight, whose amplitude depends also on galaxy and magnification biases. Hence the correlation introduced by cosmic magnification is survey dependent. The studied of this effect deserves a detailed studied which is beyond the scope of our paper.

\begin{figure}[htb!]
\begin{center}
\includegraphics[width=7cm]{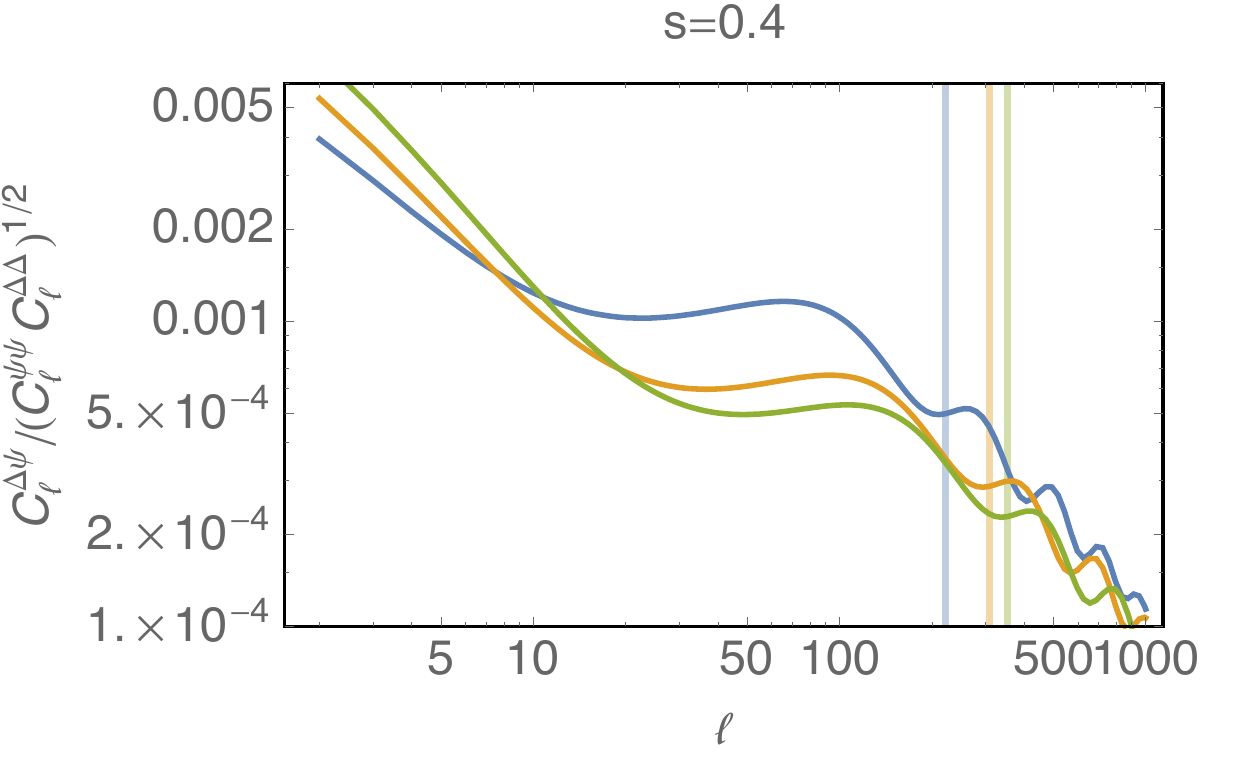}
\caption{We show the correlation between the galaxy number counts $\Delta$ and the lensing potential $\psi$ for magnification bias $s=0.4$ (galaxy bias $b=1$ and evolution bias $f_\text{evo} = 0$). Different colors refer to different redshifts: $z=2$ (blue), $z=4$ (orange), $z=6$ (green). Vertical lines show the BAO scale at the three different redshifts.
}
\label{fig:Delta_psi_s0.4}
\end{center}
\end{figure}
Even if in this work we are interested in studying the lensing effect beyond the linear cosmic magnification, we compute the amplitude of the correlation between the number counts $\Delta$ and the lensing potential $\psi$.
This correlation depends clearly on the magnification bias. Indeed as shown in Fig.~\ref{fig:Delta_psi_s0.4}, for $s=0.4$ the correlation induced by cosmic magnification vanishes and the non-vanishing correlation is induced by other integrated effects, namely ISW and time-delay. 
The smallness of this correlation for $s=0.4$ shows that the sources and the deflection angle can be considered as uncorrelated.
Interestingly, differently from Fig.~\ref{fig:delta_psi} where relativistic effects are not considered, on large scales the amplitude of the correlation increases for high redshifts on Fig.~\ref{fig:Delta_psi_s0.4}. So even if the sources and the lenses are more separated in comoving space, the sub-leading relativistic effects give a larger contribution for sources at high redshift.

\begin{figure}[htb!]
\begin{center}
\includegraphics[width=5cm]{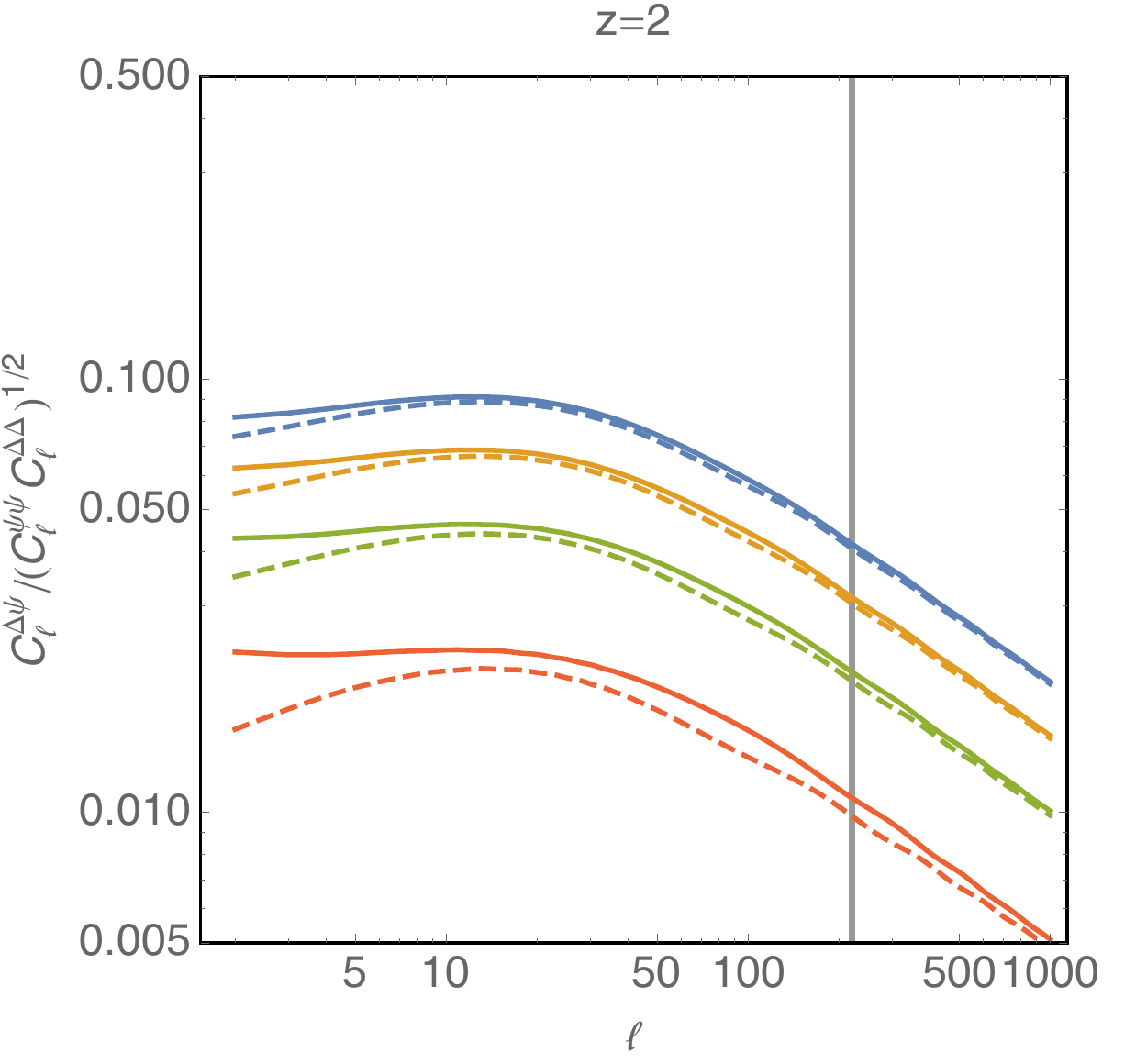}
\includegraphics[width=5cm]{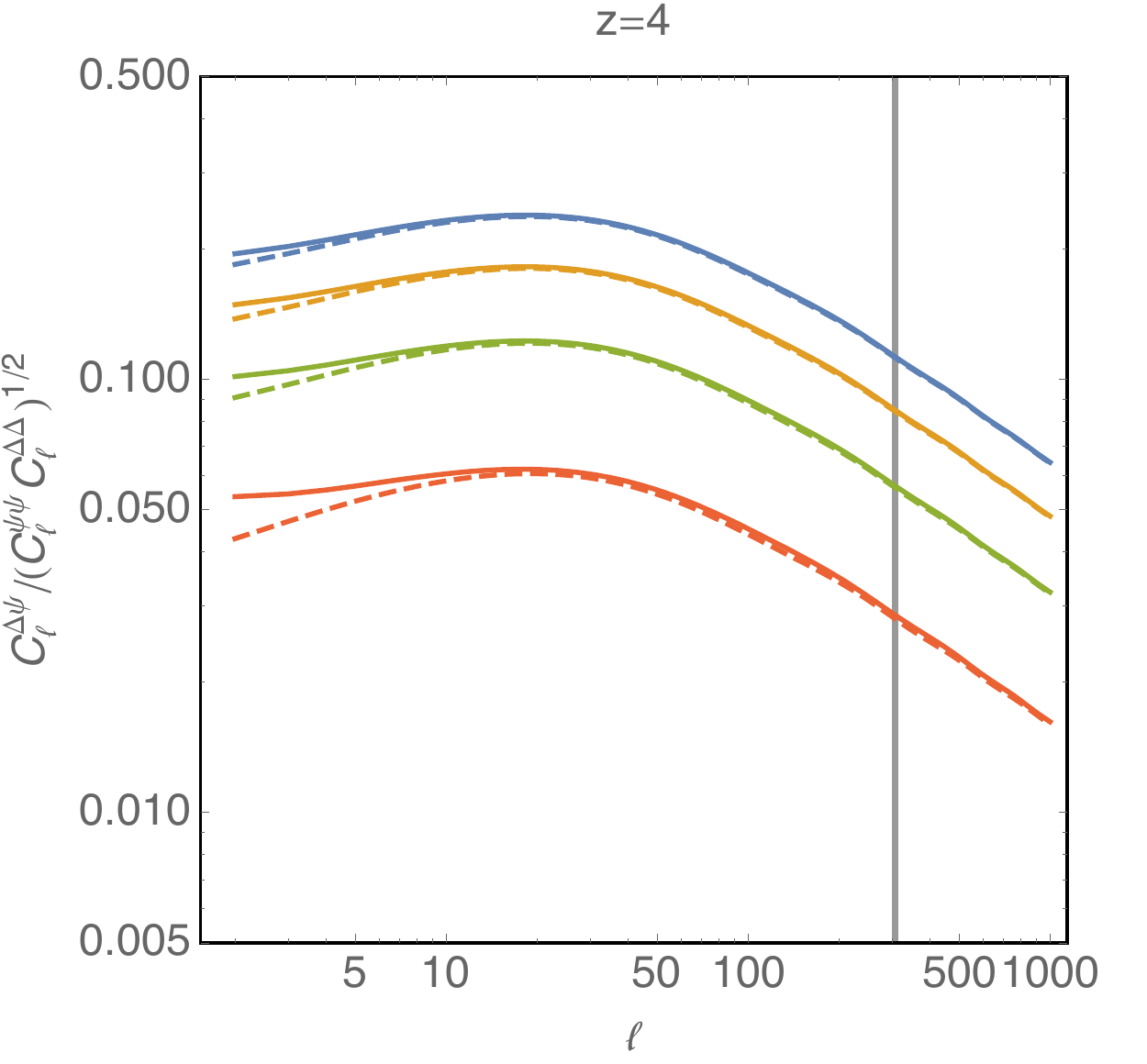}
\includegraphics[width=5cm]{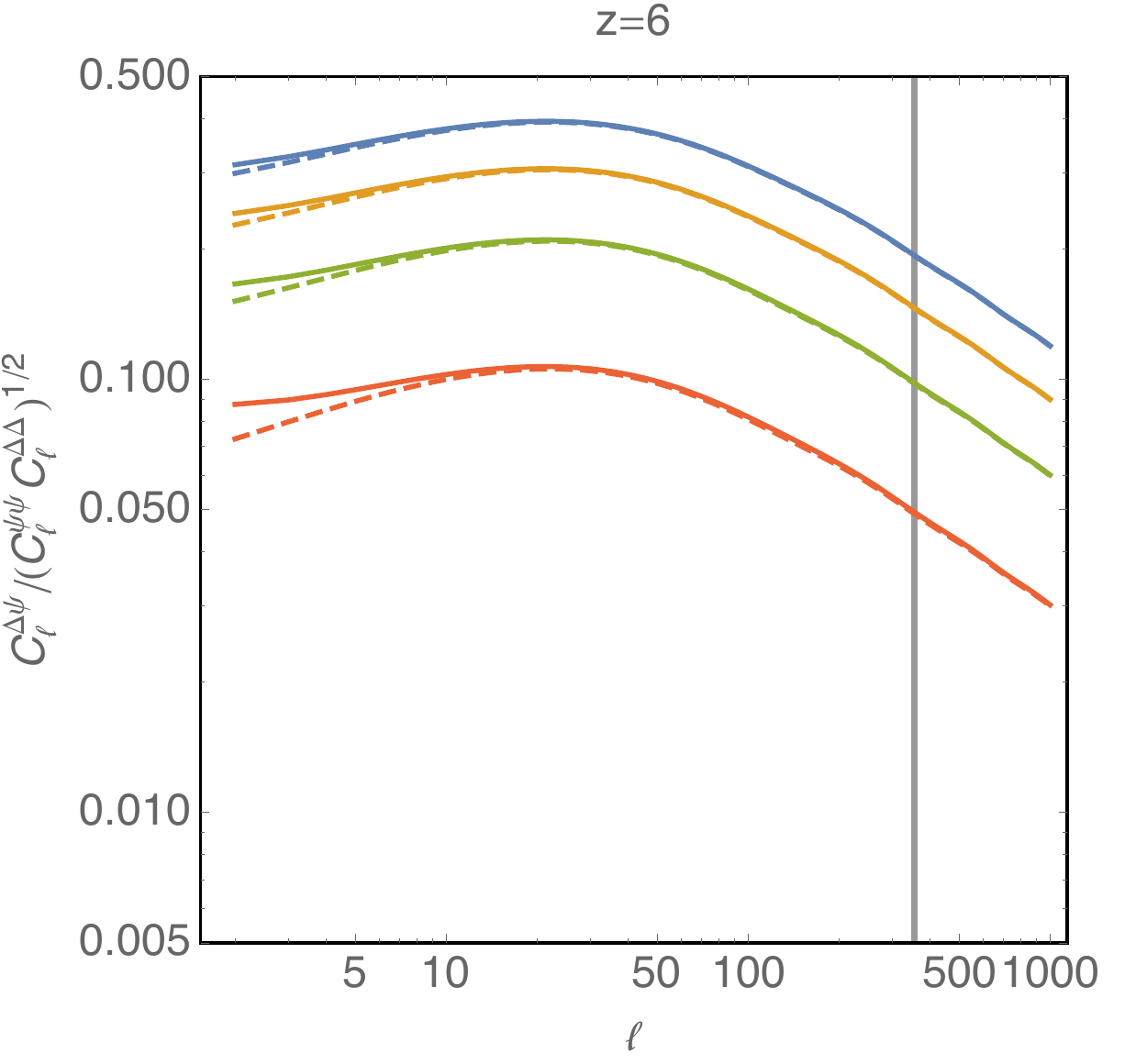}
\caption{We show the correlation between the galaxy number counts $\Delta$ and the lensing potential $\psi$ for different values of magnification bias $s$ (with galaxy bias $b=1$ and evolution bias $f_\text{evo} = 0$). Different colors refer to $s=0.3$ (red, dashed), $s=0.5$ (red, solid), $s=0.2$ (green, dashed), $s=0.6$ (green, solid), $s=0.1$ (orange, dashed), $s=0.7$ (orange, solid), $s=0$ (blue, dashed) and $s=0.8$ (blue, solid). The three panels refer to different redshifts: $z=2$, $z=4$ and $z=6$. Dashed lines indicate negative values. The gray vertical line refers to the BAO scale in angular space.
}
\label{fig:Delta_psi_s}
\end{center}
\end{figure}
As shown in Fig.~\ref{fig:Delta_psi_s} the amplitude of the correlation between the galaxy number counts $\Delta$ and the lensing potential $\psi$ strongly depends on the magnification bias $s$. Interestingly the amplitude of the correlation is about $1\%$ to $5\%$ at BAO scales for reasonable magnification bias values until $z=2$, while for higher redshifts the correlation is more relevant. Therefore we believe that our results can be approximately used also when $s \neq 0.4$, apart for sources at very high redshift. Nevertheless a more quantitative approach would require numerical simulations.
By comparing Figs.~\ref{fig:Delta_psi_s0.4} and \ref{fig:Delta_psi_s} we remark also that the main long range correlation is induced by cosmic magnification, as expected. This explains also why values of magnification bias which deviate equally (up to a sign) from $s=0.4$ (i.e.~solid and dashed lines in Fig.~\ref{fig:Delta_psi_s}) give approximately the same correlation amplitude, apart at very large scales.

Therefore, for the purpose of this derivation we work under the approximation that the deflection angle and the sources are uncorrelated, as we did for the derivation in term of the matter power spectrum in real space.

We define now the matrix
\bea
A_{\alpha \beta}  \left( \bxx_\perp \right) &\equiv& \langle \nabla_\alpha \psi \left( \bx \right) \nabla_\beta \psi \left( \bx + \bxx_\perp \right) \rangle 
\nonumber \\
&=&  \frac{1}{\left( 2 \pi \right)^2} \int d^2\ell d^2 \ell' \langle \psi \left( \bell \right) \psi \left( \bell' \right) \rangle\nabla_\alpha e^{i \bx \cdot \bell} \nabla_\beta e^{i  \left( \bx + \bxx_\perp \right)  \cdot \bell'}
\nonumber \\
&=&  \frac{1}{\left( 2 \pi \right)^2} \int d^2\ell d^2 \ell' \ell_\alpha \ell'_\beta \langle \psi \left( \bell \right) \psi \left( \bell' \right) \rangle  e^{i \bx \cdot \bell}e^{i  \left( \bx + \bxx_\perp \right)  \cdot \bell'}
\nonumber \\
&=&  \frac{1}{\left( 2 \pi \right)^2} \int d^2\ell \ \ell_\alpha \ell_\beta C^\psi_\ell  e^{- i \bxx_\perp  \cdot \bell}
\eea
where we have used the lensing potential power spectra defined in Eq.~\eqref{lens_pot_spectrum}, and dropped the redshift dependence for sake of simplicity.
Because of statistical homogeneity and isotropy, the matrix $A_{\alpha \beta}  \left( \bxx_\perp \right)$ can depend on $\bxx_\perp$ only through
\be
A_{\alpha \beta}  \left( \bxx_\perp \right) = \frac{1}{2} A_0 \left( \Delta x_\perp \right) \delta_{\alpha \beta} - A_2 \left( \Delta x_\perp \right) \left( \frac{ \bxx_{\perp \alpha} \bxx_{\perp \beta}}{\Delta x_\perp^2}  - \frac{1}{2} \delta_{\alpha \beta} \right) \, ,
\ee
where
\bea
A_0 \left( \Delta x_\perp \right)  &=&
 \frac{1}{2\pi} \int d\ell \  \ell^3 C_\ell^\psi J_0 \left( \ell \Delta x_\perp \right) \, ,
\\
A_2 \left( \Delta x_\perp \right) &=& \frac{1}{2 \pi} \int d \ell \ \ell^3  C_\ell^\psi  J_2 \left( \ell \Delta x_\perp \right) \, .
\eea
We notice how $A_0$ and $A_2$ are trivially related to $T_2$ and $D$, defined in Eq.~\eqref{eq:T2} and~\eqref{eq:D}, respectively. With these definitions, we can rewrite the expectation value as
\be
\langle \left[ \bell \cdot \left( \nabla_\perp \psi_1 - \nabla_\perp \psi_2 \right) \right]^2 \rangle =
\ell^2 \left[   A_0 \left( 0 \right) -  A_0 \left( \Delta x_\perp \right)  + A_2\left( \Delta x_\perp \right) \cos 2 \varphi \right] \, ,
\ee
where $\varphi$ denotes the polar angle on the plane orthogonal to the line of sight $\bn$.

The lensed correlation function is then given by
\bea
&&\tilde \xi \left( z_1 ,z_2 , \Delta x_\perp \right) =
\nonumber \\
&=&
\frac{1}{\left( 2 \pi \right)^2 }\int d^2 \ell C^{\Delta\Delta}_\ell \left( z_1 , z_2 \right) e^{-i \bell \cdot  \bxx_\perp } \exp \left( - \frac{\ell^2}{2} \left(A_0 \left( 0 \right) -  A_0 \left( \Delta x_\perp \right)  + A_2\left( \Delta_\perp \right) \cos 2 \varphi \right) \right) 
\nonumber \\
&\sim&
\frac{1}{\left( 2 \pi \right)^2 }\int d^2 \ell C^{\Delta\Delta}_\ell \left( z_1 , z_2 \right) e^{-i \bell \cdot  \bxx_\perp } \exp \left( - \frac{\ell^2}{2} \left(A_0 \left( 0 \right) -  A_0 \left( \Delta x_\perp \right) \right)  \right) \left(  1 - \frac{\ell^2}{2} A_2\left( \Delta x_\perp \right) \cos 2 \varphi  \right) 
\nonumber \\
&=& \frac{1}{ 2 \pi  }\int d \ell  ~ \ell ~ C^{\Delta\Delta}_\ell \left( z_1 , z_2 \right) \exp\! \left[ - \frac{\ell^2}{2} \left(A_0 \left( 0 \right) -  A_0 \left( \Delta x_\perp \right) \right)  \right] \left( J_0 \left(  \ell \Delta x_\perp \right) + \frac{\ell^2}{2} A_2 \left( \Delta_\perp \right) J_2 \left( \ell r \right) \right) \nonumber \\
\eea
and the lensed redshift dependent angular power spectra
\be
\tilde C^{\Delta \Delta}_\ell \left( z_1 , z_2 \right) = \frac{1}{4 \pi} \int d\Delta x_\perp \ \Delta x_\perp J_0 \left( \Delta x_\perp \ell \right) \tilde \xi \left( \Delta x_\perp \right) \, .
\ee
We remark that this solution fully agrees with the well-know expression for the CMB lensing, just replacing the angular power spectrum of number counts, $C^{\Delta \Delta}_\ell$, with the unlensed CMB temperature power spectrum.

\bibliographystyle{JHEP}
\bibliography{biblio_lensBAO}
\end{document}